\documentclass[10pt,journal,compsoc]{IEEEtran}
\ifCLASSOPTIONcompsoc
  \usepackage[nocompress]{cite}
\else
  \usepackage{cite}
\fi
\usepackage{amsmath,amssymb,amsfonts}
\usepackage{algorithmic}
\usepackage{graphicx}
\usepackage{textcomp}
\usepackage{xcolor}
\usepackage{balance}
\usepackage{amsmath}
\DeclareMathOperator*{\argmax}{argmax}
\usepackage{subcaption}
\def\BibTeX{{\rm B\kern-.05em{\sc i\kern-.025em b}\kern-.08em
    T\kern-.1667em\lower.7ex\hbox{E}\kern-.125emX}}
\usepackage{comment}
\usepackage{enumitem}
\usepackage{bm}
\usepackage{cleveref}
\usepackage{algorithmic}
\usepackage[ruled, linesnumbered]{algorithm2e}
\usepackage{verbatim}

\begin{document}

\title{Task Placement and Resource Allocation for Edge Machine Learning: A GNN-based Multi-Agent Reinforcement Learning Paradigm}

\author{
        Yihong~Li, Xiaoxi~Zhang,~\IEEEmembership{Member,~IEEE,}
        Tianyu~Zeng,
        Jingpu~Duan,~\IEEEmembership{Member,~IEEE,}

        Chuan~Wu,~\IEEEmembership{Senior Member,~IEEE,}
        Di~Wu,~\IEEEmembership{Senior Member,~IEEE,}
        and Xu~Chen,~\IEEEmembership{Senior Member,~IEEE}
\thanks{
Y. Li, X. Zhang, T. Zeng, D. Wu, and X. Chen are with the School of Computer Science and Engineering, Sun Yat-sen University, Guangzhou 510006, China (e-mail: \{liyh253, zengty\}@mail2.sysu.edu.cn; \{zhangxx89, wudi27, chenxu35\}@mail.sysu.edu.cn).
}
\thanks{
J. Duan is with the Department of Communications, Peng Cheng Laboratory, Shenzhen 518066, China, and also with the Institute of Future Networks, Southern University of Science and Technology, Shenzhen 518055, China (e-mail: duanjp@pcl.ac.cn).
}
\thanks{
C. Wu is with the Department of Computer Science, The University of Hong Kong, Hong Kong (e-mail: cwu@cs.hku.hk).
}
\thanks{Xiaoxi Zhang is the corresponding author.}
\thanks{A previous version~\cite{INFOCOMTapFinger} appears at INFOCOM 2023. This journal article includes new results on algorithm design and implementation, especially updating the system to be able to schedule multiple tasks simultaneously rather than selecting a single task in each timestep. System performance in task completion time and several other metrics is significantly improved.}

}

\IEEEtitleabstractindextext{%

\begin{abstract}
Machine learning (ML) tasks are one of the major workloads in today's edge computing networks. Existing edge-cloud schedulers allocate the requested amounts of resources to each task, falling short of best utilizing the limited edge resources for ML tasks. This paper proposes {\em TapFinger}, a distributed scheduler for edge clusters that minimizes the total completion time of ML tasks through co-optimizing task placement and fine-grained multi-resource allocation. To learn the tasks' uncertain resource sensitivity and enable distributed scheduling, we adopt multi-agent reinforcement learning (MARL) and propose several techniques to make it efficient, including a heterogeneous graph attention network as the MARL backbone, a tailored task selection phase in the actor network, and the integration of Bayes' theorem and masking schemes. We first implement a {\em single-task scheduling} version, which schedules at most one task each time. Then we generalize to the {\em multi-task scheduling} case, in which a sequence of tasks is scheduled simultaneously. Our design can mitigate the expanded decision space and yield fast convergence to optimal scheduling solutions. Extensive experiments using synthetic and test-bed ML task traces show that {\em TapFinger} can achieve up to 54.9\% reduction in the average task completion time and improve resource efficiency as compared to state-of-the-art schedulers.
\end{abstract}
}
\maketitle

\IEEEraisesectionheading{\section{Introduction}\label{sec:intro}}

Edge computing is a distributed computing paradigm that extends cloud capabilities to the edge for better quality of service (QoS) and data privacy protection.
By moving the service and resources closer to the end users, it can provide highly available services and reduce the service latency significantly. 
As one of the major application workloads today, edge-based machine learning (ML) applications, ranging from traffic prediction to production workflow monitoring, commonly process online data streams generated 
on the edge~\cite{edge_video2}. Due to resource limitation of edge devices, these ML tasks have been deployed in edge clusters~\cite{Ekya2021NSDI,edge_video2}, e.g., NVIDIA EGX\cite{nvidia_egx}, Microsoft Azure Edge~\cite{Azure_edge}, and AWS Outposts~\cite{aws_outposts}. 
Managed by orchestration tools, they can be running on sufficient CPUs and GPUs, as well as customized software toolkits and network interface cards, e.g., for encrypted IoT sensor data~\cite{nvidia_egx}.
However, the resources at the edge clusters are still limited. At the core of optimizing the QoS of edge-based ML applications is efficient resource utilization while learning the needed ML models timely. 

\begin{figure}[t]
  \includegraphics[width=0.87\linewidth]{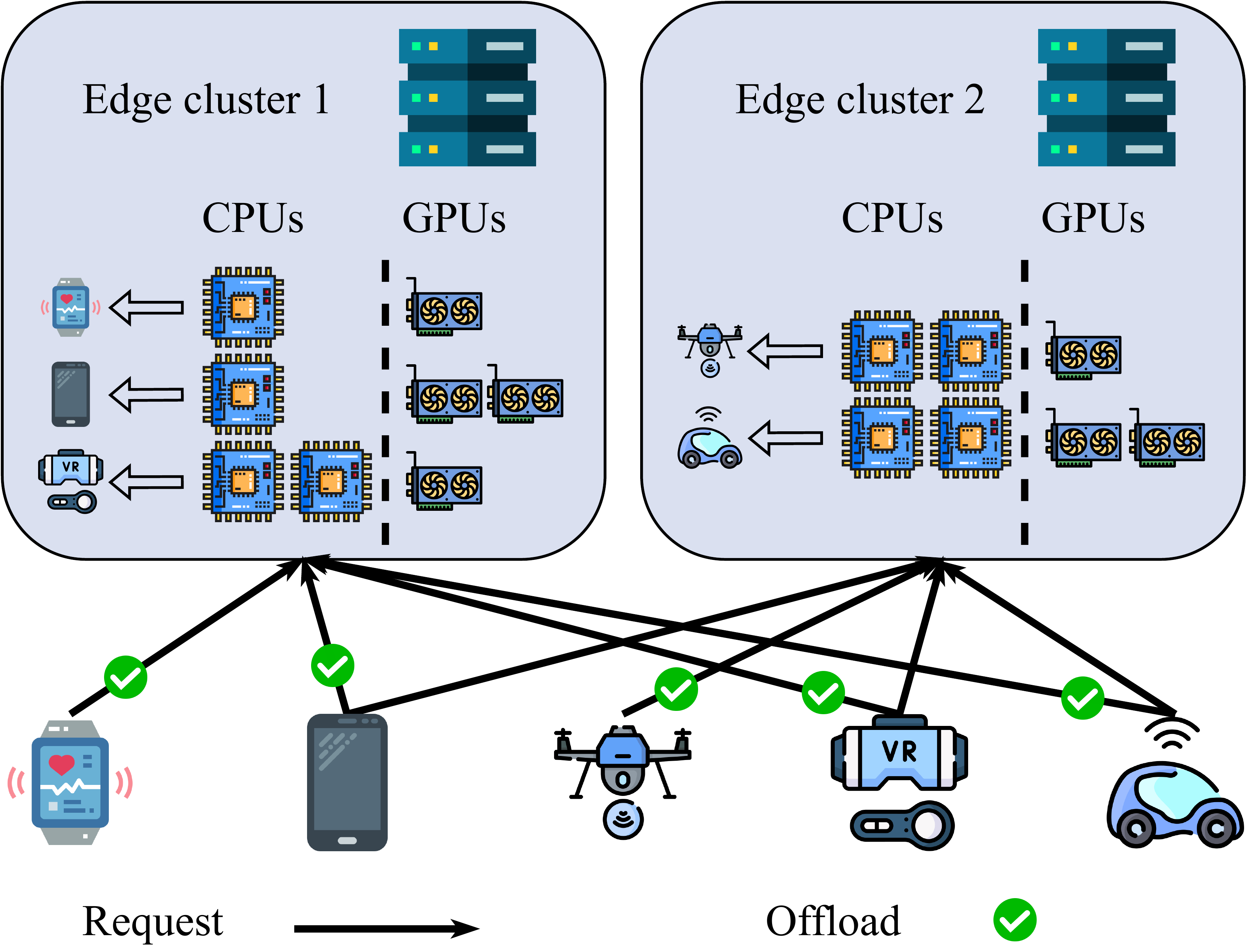}
  \centering
  \caption{Fine-grained allocation and placement for heterogeneous ML tasks in a multi-cluster edge network.}
  \label{fig:problem}
\end{figure}

With various models and datasets, ML training and inference tasks have uncertain and diverse performance~\cite{peng2018optimus,Tiresias2019NSDI,elastic2021NSDI}, making it hard to achieve optimized resource efficiency. 
Practical schedulers such as 
YARN~\cite{yarn2013}, Kubernetes~\cite{burns2016borg}, KubeEdge~\cite{kubeedge}, and OpenYurt~\cite{openyurt}
generally adopt pre-set rules for resource allocation. Those policies rely on accurate resource estimation of the tasks, while the resource demands of ML tasks are typically elastic and uncertain (e.g., the amount of time needed for model convergence), accommodating various performance-resource tradeoffs
~\cite{bao2019deep}. Learning-based cloud-edge schedulers have been proposed 
to address this uncertainty~\cite{tuli2020dynamic,peng2021dl2,bao2019deep,zhao2021large,han2021kais}. However, they 
cannot generalize to our scenario, where more complex decision dependencies need to be better encoded, e.g., for further addressing the high-dimensionality of problem inputs and decision variables.
In fact, fine-grained resource allocation and strategic task placement are exceptionally important to maximize the aggregate performance of edge ML tasks.
As illustrated in Fig.~\ref{fig:problem}, due to the data intensity and low latency requirement, mobile edge devices need to offload their ML tasks to the ``right'' edge cluster for achieving the model convergence in the minimum time. 
Besides, different resources, e.g., CPU and GPU, can affect task performance, necessitating multi-resource allocation schemes. 
Considering the complexity of dynamic network connectivity, job interference~\cite{DLworkloads2021SC}, and multi-resource contention,  
we ask: how to design a scalable, fine-grained, and far-sighted resource scheduler customized for edge ML? 

To realize this, we propose {\em TapFinger}, a distributed scheduler to {\em jointly optimize task placement and fine-grained multi-resource allocation across edge clusters}, with a goal of minimizing total completion time of ML tasks. To achieve this, we face the following fundamental challenges. 

\textbf{Fine-grained 
resource allocation.}
Different from existing edge-cloud schedulers which allocate the requested resource amounts to each task~\cite{tuli2020dynamic,Tiresias2019NSDI}, 
fine-grained resource provisioning~\cite{elastic2021NSDI} according to demand and supply can achieve better resource efficiency and application QoS.
The challenge is then on the side of edge schedulers, 
in 
predicting the performance of each concurrently running task and strategically choosing the best resource amounts from a huge solution space for maximizing the aggregate task performance, for instance, the required training time to achieve a certain accuracy.

\textbf{Uncertain impact of different resources}. Most ML task schedulers allocate a single type of resource, e.g., GPU, in the cloud~\cite{Gandiva2018OSDI,zhao2021large,elastic2021NSDI} or edge settings~\cite{Ekya2021NSDI}, failing to capture the effects of multiple types of resources on the task performance. 
A multi-resource allocation problem is typically NP-hard even with perfect knowledge of the problem input~\cite{multiresource2013infocom}, which is more challenging when the task performance is unknown given the resources.
Reinforcement learning (RL) can efficiently deal with uncertainty~\cite{Sutton2005ReinforcementLA,mirhoseini2017device,mao2016resource} but is still under-explored for online combinatorial problems with complex constraints. 

\textbf{Distributed scheduling across edge clusters.}
Practical edge-cloud clusters are usually managed by a central coordinator~\cite{burns2016borg,yarn2013}. 
However, centralized scheduling processes the global information over multiple geographical regions and often suffers from poor scalability. 
%
A decentralized approach is therefore preferable to reduce the decision space and can enable better system reliability. Nonetheless, decentralized scheduling may lead to sub-optimal task performance if 
self-optimizing decisions in each edge cluster independently.
Effective interactions among the distributed schedulers are crucial for optimizing global resource efficiency in the entire edge network.



To address these challenges, {\em TapFinger} adopts a multi-agent reinforcement learning (MARL) approach. Compared to rule-based distributed optimization, it can better generalize to unseen task characteristics and different edge networks. To achieve efficient MARL for our distributed resource scheduling, we make the following technical contributions:




{\em First}, we propose a state abstraction technique to enable efficient information interactions among distributed agents. Conventional MARL approaches can be sub-optimal due to the partial observability of each agent. 
%
We use a heterogeneous graph attention network (HAN)~\cite{wang2019heterogeneous} to encode rich semantic information of different edge components and their 
dependencies. The local observations are then passed as messages among the agents and provide them with a global view of the environment. 
The raw states are then mapped into a compressed space and efficiently improve the learning ability of the agents.

{\em Second,} we propose several techniques to augment our MARL actor network and the training method. 
To decompose our actions on task placement and resource allocation, we design a task selection phase via a pointer network module~\cite{vinyals2015pointer,kool2018attention}, inspired by natural language processing (NLP) techniques. We then construct a conflict resolution module based on Bayes' theorem to coordinate different agents' decisions. In addition, we use masking schemes to encode our constraints and filter out the gradient propagation that is irrelevant to the conflicts, which effectively speed up the MARL convergence. By separating the real-world timestep and the RL timestep, we further extend the framework from that every edge cluster makes the scheduling decision for only one task, to simultaneously allocating the available resources to multiple tasks, in each real-world interval.

{\em Third,} we conduct extensive experiments and ablation analysis on synthetic and test-bed ML task traces. We observe a significant reduction in terms of the average task completion time, compared to representative scheduling algorithms. The experiments also show that our algorithm can capture the diverse resource sensitivities of different types of ML tasks and effectively improve resource efficiency. {\em TapFinger} also demonstrates its scalability as it maintains significant superiority with varying numbers of edge clusters, system span, and task arrival rates.

\section{Related Work}\label{sec:related}
\textbf{Resource scheduling for DL workloads.}
Although general-purpose task scheduling algorithms such as Dominant Resource Fairness~\cite{ghodsi2011drf}, 
Shortest Remaining Job First (SRJF)~\cite{srtf2016OSDI},
Tetris~\cite{grandl2014tetris}, and their improved variants~\cite{wang2016tsf,chowdhury2016hug,khamse2018efficient} have been extensively studied, the strategies tailored to machine learning (ML) workloads remain premature~\cite{gao2022deep}. 
Several pioneering scheduling frameworks capture the uncertain execution times of DL tasks and aim to minimize the completion time.
Peng et al.~\cite{peng2018optimus} design Optimus to schedule distributed DL training tasks by predicting their training speed as a function of allocated resources for each task. The scheduler can then incrementally adjust the numbers of parameter servers and workers for all tasks. 
To exploit the cyclic patterns in DL tasks, Xiao et al.~\cite{Gandiva2018OSDI} 
develop primitives based on existing DL frameworks to
realize efficient time-slicing and profiling-driven introspection. Tiresias~\cite{Tiresias2019NSDI} combines least-attained service scheduling and multilevel feedback queues to design a preemptive scheduling algorithm. 
In \cite{elastic2021NSDI}, the Apathetic Future Share algorithm is proposed to achieve efficient resource sharing, with a dedicated elastic-share DL training system implemented to validate the advantages of the proposed algorithm. 
A few recent works achieve elastic GPU cluster scheduling for DL tasks using
multi-resource pipelining and inter-job interleaving~\cite{peng2019generic,zhao2022multi},
novel profiling methods~\cite{Ekya2021NSDI,jajoo2022case} or new performance metrics~\cite{pullux2021OSDI,elastic2021NSDI}.
Wang et al.~\cite{wang2022preemptive} propose an online preemptive ML task scheduling algorithm for edge-cloud networks.
In general, these are rule-based algorithms that rely on accurate estimations of task characteristics. This paper, instead, aims to enable a self-optimization framework without hand-crafting prediction models for capturing uncertain task performance.

\textbf{Task scheduling with deep reinforcement learning.}
Deep Reinforcement learning (DRL) has demonstrated its effectiveness in online decision making including task scheduling and resource allocation.
Mao et al.~\cite{mao2016resource} introduces DRL to solve general multi-resource allocation in computer systems.
Several works~\cite{mirhoseini2017device,cheong2019scarl,lee2020panda} leverage the pointer network mechanism~\cite{vinyals2015pointer} for task placement. 
However, they have not considered ML workloads or fine-grained resource allocation. 
Wang et al.~\cite{wang2020job} propose an ML feature-based task scheduling system with superiority in both job completion time and training accuracy.
They use the trace data from a designed heuristic scheduling method to train a deep reinforcement learning model. 
In~\cite{bao2019deep}, a DRL method that fuses rule-based policies is proposed for ML task placement. Utilizing offline supervised learning and online reinforcement learning, Peng et al.~\cite{peng2021dl2} propose a trace-driven task scheduler for DL clusters 
using offline supervised learning to boost the DRL training.
Recent works have also extended DRL for resource allocation in edge clusters. Tuli et al.~\cite{tuli2020dynamic} utilize asynchronous advantage actor-critic to decide the allocation of edge-cloud resources by regarding the distributed resources as a large number of independent hosts. Different tasks and hosts are divided among multiple agents for DRL training. At each step, an agent makes decisions about scheduling new tasks and migrating old tasks, and it outputs a priority list of hosts for each task.
These works adopt single-agent DRL and do not consider fine-grained resource allocation for ML tasks or jointly optimize task placement and resource allocation.

\textbf{GNN-based DRL methods for task scheduling.}
Beyond applying DRL approaches with standard formulation of the environmental states and actions, there is a trend of designing state abstraction to boost DRL training. Graph neural network (GNN) is a promising model that can encode the dependencies across state features. We have identified that two recent works are similar to ours methodology-wise. First, Zhao et al.~\cite{zhao2021large} adopt GNN to encode cluster topology and server configurations, motivated by the interference of co-located DL tasks. 
They design an MARL algorithm for scheduling among GPU clusters. Each agent decides either to serve the task or to forward it to other agents.
Second, Han et al.~\cite{han2021kais} propose a two-timescale scheduler for Kubernetes-oriented edge-cloud clusters. They introduce GNN-base DRL for request dispatch and MARL 
scheduling for service orchestration.
However, the above works do not consider instant fine-grained resource allocation, hence being
detrimental to resource efficiency.

\section{System Model and Problem Definition}\label{sec:problem}

In what follows, we describe our system infrastructure first, and then formulate the task placement and resource allocation problem. Key notations are summarized in Table~\ref{table:notations}. We define $[X]\triangleq\{1,2,\cdots, X\}$ in our description. 

\subsection{Edge Clusters}\label{subsec:cluster}
We consider that $N$ geo-distributed edge clusters provide low-latency machine learning services to local edge devices. Today's edge clusters are equipped with sufficient computational ability to train and run deep neural network (NN) models\cite{nvidia_egx,Azure_edge}. 
For compliance requirements and restrictive data policies, ML training and inference tasks may be required to run on certified edge clusters.
 We consider a total of $R$ types of computation resources, e.g., CPU, GPU, memory, and bandwidth, offered by each edge cluster to perform ML tasks. The clusters are not necessarily homogeneous in real-world settings. 
Each resource $r\in [R]$ has a distinct smallest unit $\delta_r$ that can be allocated, and the resource capacity $C_{r,n}$ is equal to the total amount of $\delta_r$ available in cluster $n$. The full set of possible discrete values that can be chosen to provision for a task is then $\{0,1, ..., C_{r,n}-1\}, \forall r\in[R], n\in [N]$. 
The tasks that come online may occupy resources for various task durations. 
Therefore, the maximum amount of each type-$r$ resource at $n$ that can be allocated at timestep $t$, defined to be $B_{r,n,t}$, can be time-varying. 

Edge computing networks are distributed in nature, 
and provides basic data exchange functions via low-cost network connections among edge clusters~\cite{xu2021imc}, unlike the stable network condition and sufficient bandwidth in cloud datacenters. It is then impracticable to conduct a large number of distributed parallel training and inference processes across multiple edge clusters~\cite{MLSYS2020_net_overhead}. Therefore, we consider in-cluster GPU-level parallel training and inference
for better QoS of user applications. Further, to fully utilize the resources of the entire network, we allow 
edge devices to 
send offloading requests to all accessible edge clusters, as in Fig.~\ref{fig:problem}. Based on the current resource availability and predictions of the future performance gain, the system will then decide which cluster to accept the request for any given task.


\begin{table}[!t]
\caption{Notation used in our formulation.}
\resizebox{\linewidth}{!}{%
\begin{tabular}{|c|c|}
\hline
$r\in [R]$          & index of resource                                           \\ \hline
$\delta_r$          & distinct smallest unit of $r$                               \\ \hline
$n\in [N]$          & index of edge cluster                                       \\ \hline
$t\in [T]$          & index of real-world timestep                                \\ \hline
$C_{r,n}$           & total capacity of type-$r$ resource in $n$                  \\ \hline
$B_{r,n,t}$         & amount of type-$r$ resource at $n$ that is available at $t$ \\ \hline
$j\in [J]$          & index of task                                               \\ \hline
$\mathcal{D}_j$     & set of edge clusters that $j$ can be offloaded to           \\ \hline
$\mathcal{Q}_{n,t}$ & set of tasks in the queue of cluster $n$ at $t$             \\ \hline
$b_{j,r}$           & minimum amount of type-$r$ resource required by task $j$    \\ \hline
$y_{j,n,t}$         & $j$ is scheduled by $n$ in $t$ (=1) or not (=0)             \\ \hline
$x_{j,n}$           & amount of type-$r$ resource in $n$ allocated to $j$         \\ \hline
$\hat{t}_j$         & arrival time of $j$                                         \\ \hline
$t_j^{\ast}$        & start time of execution of $j$                              \\ \hline
$\phi_j(\cdot)$         & completion time of $j$                                  \\ \hline
\end{tabular}%
}
\label{table:notations}
\end{table}

\subsection{Timesteps, Task Arrivals, and Placement}\label{subsec:task_selection}
Since we aim to optimize the aggregate 
QoS over any exogenously determined 
workload, 
we consider a total of $T$ timesteps, each of which can be non-uniform and event-driven, e.g., by the arrival, the execution, or the completion of tasks.
Note that we divide all the events into different timesteps even if they happen at the same time. For example, one of the edge clusters is making multiple scheduling decisions simultaneously, and the interval between these decisions can be zero or infinitely small. Defining such logical timesteps in our model is to reflect the causal relationships between scheduling decisions across multiple tasks in each real-world time, i.e., the {\em scheduling order}, and enable more scalable optimization methods compared to inferring a vector of scheduling decisions simultaneously at each timestep.
Within the $T$ timesteps, a total of $J$ 
ML tasks arrive one by one, with ties broken by associating their arrival times with distinct timesteps $t\in [T]$ (a batch of tasks that come at the same real-world time are randomly assigned to different logical timesteps).

The differences in implementing such logical timesteps between our two scenarios are: (i) in the {\em single-task scheduling} case, the timesteps, in which the edge schedulers make decisions correspond to evenly split real-world time points; (ii) in the {\em multi-task scheduling} case, the timesteps can associate with scheduling multiple tasks to execute in the same real-world time point, or times with infinitely small intervals.  
We emphasize that, in both cases, multiple tasks may share the resources and execute concurrently in each edge cluster, but more than one task can be scheduled in each real-world time point in the {\em multi-task scheduling} case, compared to a single task in the {\em single-task scheduling} case. As will be described in the following section \ref{subsec:resource_allocation}, resource allocation decision is also made for at most one task in any given logical timestep $t$, covering both of the aforementioned scenarios. More differences between these two scenarios are in the implementation of the algorithms.

Heterogeneous ML tasks can, for instance, include image classification training tasks over ConvNets, speech recognition inference tasks using transformers, etc.
To take advantage of edge computing technologies such as offloading, we consider that each ML task generated locally on resource-scarce edge devices needs to be offloaded to an edge cluster. Due to network limitations, the edge devices may only connect to a subset of the edge clusters. We define $\mathcal{D}_j$ to be the set of eligible edge clusters that task $j$ can be offloaded to and $\mathcal{Q}_{n,t}$ to be the set of tasks in the queue of cluster $n$ at $t$. We allow edge devices to send requests to all the edge clusters in $\mathcal{D}_j$ simultaneously
to avoid the long wait for being scheduled, but restrict that one task can only be finally offloaded to one edge cluster, as shown in Fig.~\ref{fig:system}. For now, we assume that the edge clusters will negotiate with each other so that only one edge server will serve the task in the end. The introduction of the specific mechanism to realize this negotiation will be deferred to Section \ref{subsec:decomposition}.

The task requests continuously arrive from the edge devices into the queue of each connectable edge cluster. The scheduler of each cluster $n$ then independently decides which task $j\in\mathcal{Q}_{n,t}$ to serve when $t$ starts. Since multiple clusters may choose the same task at the same time, a coordinator is needed to choose one of the clusters to serve the task. We define a binary variable $y_{j,n,t}$ to denote whether $j$ is scheduled by cluster $n$ in timestep $t$, and require $\sum_{j\in\mathcal{Q}_{n,t}}y_{j,n,t}\le 1$, i.e., each task $j$ is only scheduled to at most one cluster at each timestep $t$. 
For each task $j$, $y_{j,n,t}$ is only positive ($=1$) at the timestep when $j$ is scheduled by $n$ and remains zero in other timesteps, indicating that {\em task $j$ will be processed by cluster $n$ until it completes}. This definition of $y_{j,n,t}$ yields a simpler way to model the resource allocation problem considering other optimization constraints, as we will elaborate below in \eqref{eq:optimization_problem}--\eqref{eq:st_x_region}.
%

Note that in modern container orchestration tools like Kubernetes, the scheduler first collects all feasible nodes in the cluster. A node may be a virtual or physical worker machine. Then the scheduler scores these node candidates according to a series of factors, including hardware/software constraints, data locality requirements, etc., and finally binds the node with the highest score to the pod~\cite{burns2016borg}, which is a set of running containers managed by the scheduler. Multi-thread scheduling can schedule tasks concurrently but still needs a transaction commit process. In essence, this is a sequential operation for one scheduler. Therefore, it is consistent with the setting of our timesteps and decision-making, where each edge cluster only schedules at most one task at any logical timestep $t$. 

\subsection{Multi-Resource Allocation}\label{subsec:resource_allocation}
\begin{figure}[t]
    \centering
    \begin{subfigure}{0.5\linewidth}
    \includegraphics[width=\linewidth]{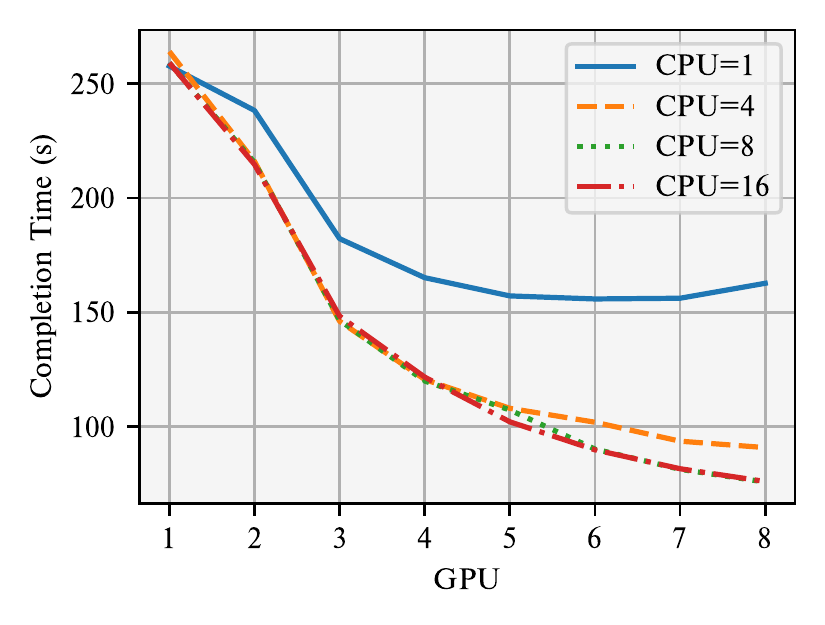}
    \caption{Varying with GPU.}
    \label{fig:CPU_on_GPU}
    \end{subfigure}\hfill
    \begin{subfigure}{0.5\linewidth}
    \includegraphics[width=\linewidth]{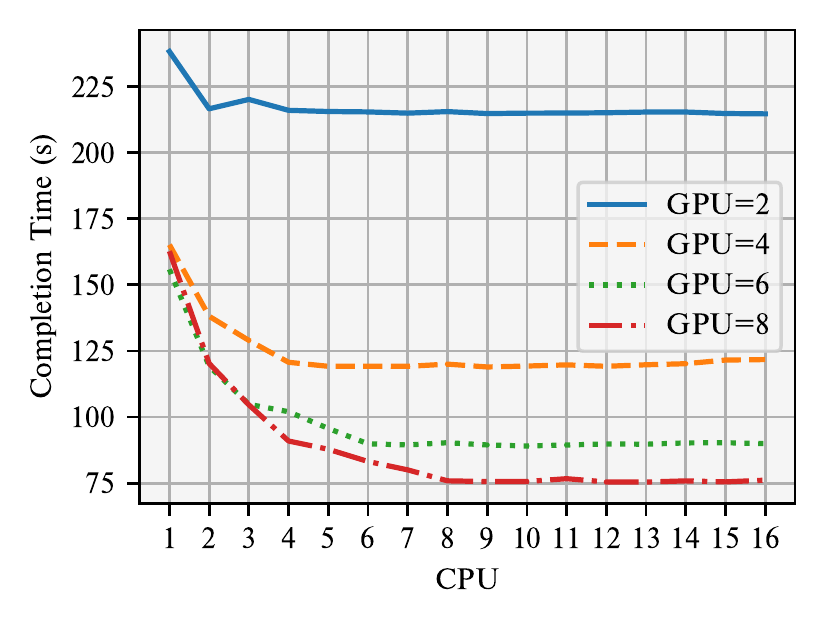}
    \caption{Varying with CPU cores.}
    \label{fig:GPU_on_CPU}
    \end{subfigure}\hfill
    \caption{Testbed results on the average completion time of training a language modeling task using a transformer network~\cite{vaswani2017attention} by varying the combination of allocated resources.}
    \label{fig:motivation}
\end{figure}

Once selecting a task $j\in \mathcal{Q}_{n,t}$, the scheduler needs to decide its resource allocation. As shown in Fig.~\ref{fig:motivation}, we identify that the completion time of a representative ML task has various sensitivities under different combinations of CPU and GPU resources. For instance, 
the completion time first 
decreases with the number of CPU cores and then becomes rather flat. Improper resource allocation, e.g., 1 CPU and 8 GPUs, may lead to a rebound in the completion time because a CPU bottleneck occurs, when the number of GPUs increases.
Motivated by our observation, 
we consider that the scheduler in each cluster decides the allocation of each type of resource. 
We define an integer decision variable $x_{j,n}$ to represent the amount of type-$r$ resource in cluster $n$ allocated to $j$. 
 Let $\hat{t}_j$, $t_j^{\ast}$, and $\phi_j$ denote the arrival time, start time of execution, and completion time of task $j$, respectively, where $t_j^{\ast}$ is decided by our scheduling decisions and $\phi_j$ is determined by not only $\hat{t}_j$ but also $t_j^{\ast}$ and our resource allocation. Since the tasks may arrive at different times and occupy the resources for various periods of time, we have a time-coupling capacity constraint $\sum_{j: t\in [t_j^{\ast}, \hat{t}_j+\phi_j)} x_{j,r,n}\le C_{r,n}$ for each $r$, $n$, and $t$. In addition, we define $b_{j,r}$ as the minimum amount of type-$r$ resource required by task $j$. Thus, each $x_{j,r,n}$ has to be at least $b_{j,r}$ if positive.

Here we assume that our resource allocation cannot be modified once decided. Although preemptive schedulers may achieve better resource utilization and reduce the average completion time~\cite{Tiresias2019NSDI,elastic2021NSDI}, they are not widely implemented in real-world ML clusters~\cite{xiao2020antman}. One of the main reasons is that pausing and resuming ML tasks using the checkpointing mechanism frequently incurs overhead, increases system complexity, and introduces potential failures, especially for training tasks~\cite{checkNrun2022nsdi}. Distributed DL training frameworks like PyTorch and Horovod provide elastic training functions that allow users to scale the number of workers during the training process. However, that might need dynamically adjusted hyper-parameters, e.g., learning rate and batch sizes~\cite{pullux2021OSDI}, which 
requires extra coding from task owners and still incurs unpredictable system overhead. These uncertainties consequently prevent reproducing the training convergence results and contradict the original intention of using edge computing to guarantee the QoS of ML tasks.

We now formalize our joint task placement and multi-resource allocation problem. The goal is to minimize the total completion time of all tasks that arrive in $[T]$, 
while satisfying the resource capacity constraints and the tasks' resource requirements. 
In practice, a training task is completed when a certain accuracy or convergence is met. The completion time of an inference task is also hard to predict due to network instability and fluctuation in available resources\cite{Fang2019edge_inference}. Therefore, the completion time of each task ($\phi_j(\cdot)$) is unknown a priori and affected by our task placement $\mathbf{y}$ and resource allocation $\mathbf{x}$ decisions for all the tasks that co-exist with $j$.
Finally, we formulate the optimization problem as follows.
\begin{align}\label{eq:optimization_problem}
\underset{\mathbf{x},\mathbf{y}}{\text{Minimize}}
  &\hspace{3mm}\sum\limits^{J-1}_{j=0}{\phi_j(\mathbf{x}, \mathbf{y})}&\\
\text{S.t.} 
 & \sum_{j\in\mathcal{Q}_{n,t}}y_{j,n,t}\le 1, \quad \forall n, t \label{eq:st_task_select}\\
 & \sum_{n\in \mathcal{D}_j}\sum_{t\in[T]} y_{j,n,t}\le 1, \quad\forall j\label{eq:st_task_placement}\\
 &y_{j,n,t}\in \{0, 1\},\quad \forall j, n, t\label{eq:st_y_region}\\
 & \sum_{j: t\in [t_j^{\ast}, \hat{t}_j+\phi_j)} x_{j,r,n}\le C_{r,n}, \quad\forall r,n,t\label{eq:st_capacity}\\
 & t_j^{\ast} = \text{max}_{n,t}\left(t\cdot y_{j,n,t}\right), \quad\forall j\label{eq:st_start_time}\\
 & x_{j,r,n} \le \text{max}_{t} \left(y_{j,n,t}C_{r,n}\right), \quad\forall j,r,n\label{eq:st_x_y}\\
 & x_{j,r,n}\in \{0\} \cup \{\mathbb{Z}\cap [b_{j,r}, C_{r,n}]\}, \forall j,r,n\label{eq:st_x_region}
\end{align}
Here, since the task scheduling decision $y_{j,n,t}$ is binary (constraint \eqref{eq:st_y_region}), \eqref{eq:st_task_select} ensures that at most one task can be added from the queue onto each cluster $n$ at each timestep $t$. 
For each task $j$, $y_{j,n,t}$ is only positive ($=1$) at the timestep when $j$ is scheduled by $n$ and remains zero in other timesteps, indicating that {\em task $j$ will be processed by cluster $n$ until it completes} (inequality \eqref{eq:st_task_placement}). Our definition of $y_{j,n,t}$ requires that our placement decision for task $j$ cannot be modified and yields a simpler way to model the resource allocation problem, although other modeling choices can also work.
Further, \eqref{eq:st_capacity} requires that each type of resource occupied by all the alive tasks $\{j \big|t\in [t_j^{\ast}, \hat{t}_j+\phi_j)\}$ in any time $t$ cannot exceed the corresponding capacity. Equation \eqref{eq:st_start_time} defines the start time that $j$ is scheduled, which is the only time that $y_{j,n,t}$ is positive for task $j$. Finally, \eqref{eq:st_x_y} and \eqref{eq:st_x_region} define the time-varying feasible region of $\mathbf{x}$ at each timestep $t$. 

Our formulation in \eqref{eq:optimization_problem}--\eqref{eq:st_x_region} mathematically models the dependencies between the decisions of task placement and resource allocation across co-existing tasks. It provides insights to our algorithm design, e.g., its impact factors that need to be observed as environmental states, rewards, and the feasible region of the actions. However, this formulation captures a centralized optimization and needs to be factorized for distributed scheduling based on MARL, which we will elaborate in Section \ref{sec:algorithm}. 

\section{Algorithm Design}\label{sec:algorithm}

\begin{figure*}
  \begin{center}
  \includegraphics[width=1\linewidth]{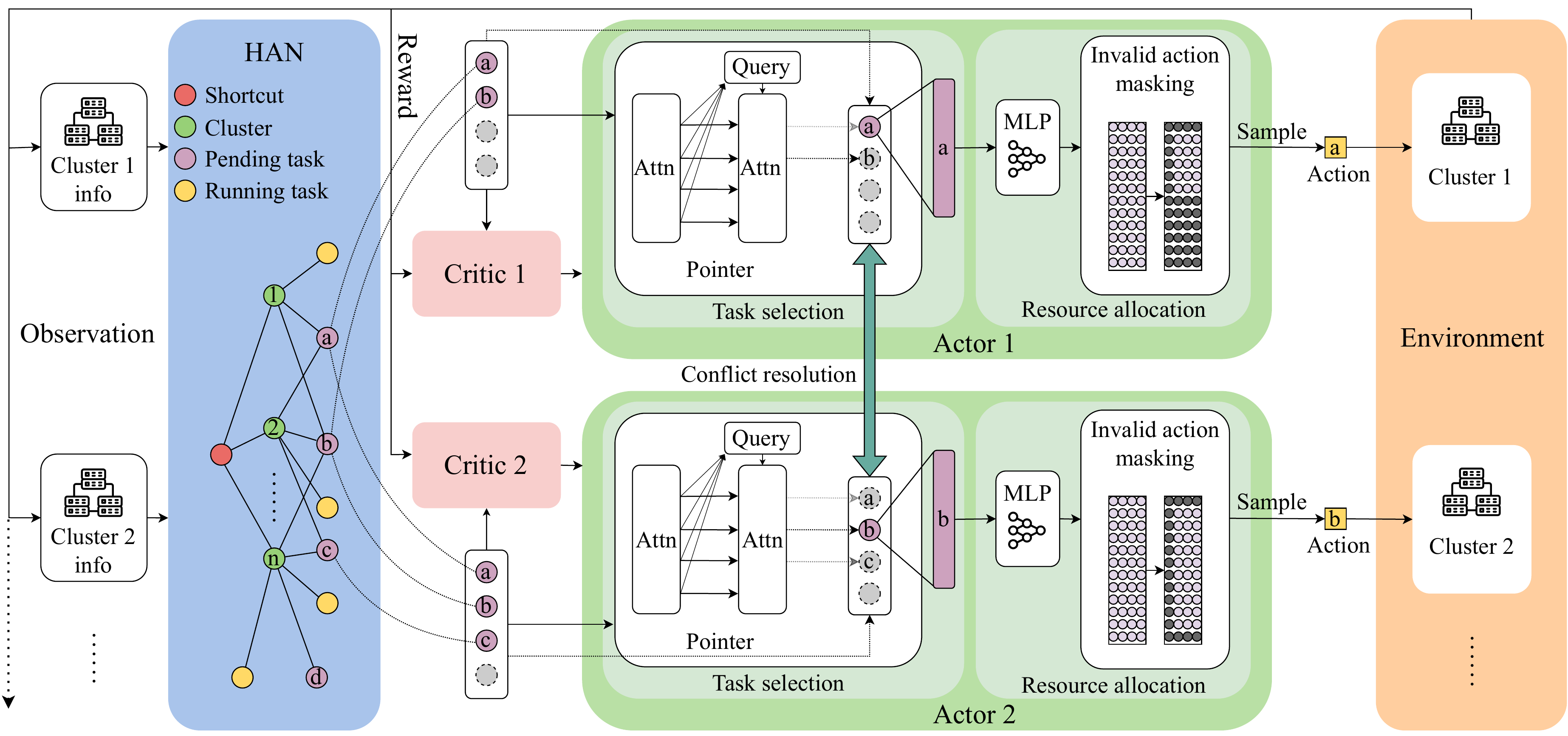}
  \caption{The design components of {\em TapFinger}.}
  \label{fig:system}
  \end{center}
\end{figure*}

We next walk through the designs of our proposed algorithm that jointly optimizes the task placement and resource allocation for ML tasks across multiple edge clusters.

To ease the presentation, we first illustrate the components of our system {\em TapFinger} for the {\bf\em single-task scheduling} scenario described in Section \ref{subsec:task_selection}, in which each agent only schedules one task at the beginning of each timestep, which is uniformly apart in the real-world time horizon. Then we discuss the extension to the {\bf \em multi-task scheduling} version in Section \ref{subsec:multitask}. Both versions of {\em TapFinger} share the same structure of learning models and algorithm components shown in Fig.~\ref{fig:system}. They only differ in the reward design and the way how the agents interact with the environment.

\subsection{Algorithm Overview}
Driven by the complexity of co-optimizing the decision variables in our offline optimization \eqref{eq:optimization_problem}--\eqref{eq:st_x_region}, we propose to enable a distributed optimization where $\mathbf{x}$ and $\mathbf{y}$ can be first independently and preliminarily decided by each cluster, and then coordinated by a central coordinator residing in one of the edge clusters. 
The central coordinator is designed for passing moderate environmental observation between different edge clusters (see Section \ref{subsec:representation}) and deciding the final scheduling decisions by resolving conflicts among clusters (see Section \ref{subsec:decomposition}). It can be replicated in multiple edge clusters in order to allow alternative replicas to take over in case the current central coordinator fails. Maintaining strong consistency at inference time is not necessary since all the states maintained in the central coordinator are volatile except for the stored model parameters.
In addition to the distributed architecture, another intuition is that the unknown completion time $\phi_j(\mathbf{x}, \mathbf{y})$ can be optimized through trials and errors, if we can learn the statistical patterns of tasks' resource sensitivities and their dependencies. These design perspectives fit with the basic idea of MARL naturally. Our goal is then to push the limit of MARL towards solving dependent decisions under time-varying constraints, which drives the design of {\em TapFinger}.
Fig.~\ref{fig:system} shows our system components, illustrating the task placement and multi-resource allocation based on the actor-critic architecture~\cite{Sutton2005ReinforcementLA} across at least two edge clusters. We introduce several techniques to the MARL algorithm, e.g., state representation through GNN, action decomposition, and loss function design based on masking schemes. We brief the concepts of state, action, and reward in this section. The technical specifics of our design are elaborated in Sections \ref{subsec:representation} and \ref{subsec:decomposition}.

\textbf{State space.}
Intuitively, the dynamics of task arrivals, connectivity between devices and clusters, as well as resource sensitivity of each task will all affect our objective function and constraints in \eqref{eq:st_capacity}--\eqref{eq:st_x_region}. 
The challenge is that it is inefficient either to stack all these factors across the entire edge network into a global state matrix or to split the state space into sub-spaces for each individual agent.
We adopt a heterogeneous graph attention (HAN) network~\cite{wang2019heterogeneous} to encode our features and their interrelation 
semantics. The output of HAN serves as the abstracted states for each agent, which compress the raw states and deliver learnable global environmental patterns.   
This design will be detailed in Section \ref{subsec:representation}.


\textbf{Action space.}
Even using MARL, the action space of each agent will be huge if directly concatenating our decisions $\mathbf{x}_n$ and $\mathbf{y}_n$ as a vector with a dimension of $(|Q_{n,t}|\times \prod_{r=1}^{R}{C_{r,n}})$. To reduce this space, we design an action decomposition technique that combines the pointer network and decision conflict resolution. 
Each agent $n$ takes the HAN embedding of the pending tasks corresponding to cluster $n$ as the input of its actor network, and outputs both the task selection and resource allocation actions. For task selection, we use transformer layers to further encode the HAN embedding and introduce a pointer mechanism~\cite{vinyals2015pointer} to decode the task selection actions. Then a conflict resolution module takes control to resolve task selection conflicts. Finally, the HAN embedding of the selected task is input into our invalid action masking module which encodes the constraints and outputs the final resource allocation. 
Technical details are shown in Section \ref{subsec:decomposition}.

\textbf{Reward.}
Note that the total completion time equals the sum of the total number of alive tasks over all timesteps, i.e., $\sum_{j}\phi_j(\mathbf{x}, \mathbf{y}) = \sum^T_t{R_t(\mathbf{x}, \mathbf{y})}$. Therefore, $R_t(\mathbf{x}, \mathbf{y})$ is equal to the total number of all tasks in the system in $t$ including the ones that are waiting or running at any edge cluster. 
We define $r_{n,t}=-R_{n,t}$ as the reward for $n$ in timestep $t$. Similar to $R_t$, $R_{n,t}$ is the number of tasks associated with the edge cluster $n$ in timestep $t$. 

\subsection{State Representation via GNN Design} \label{subsec:representation}
Reviewing the structure of our optimization problem, a good scheduler should account for the workload on each cluster and their currently allowed allocation choices, i.e., the minimum required and maximum available resource amounts based on \eqref{eq:st_capacity}--\eqref{eq:st_x_region}. It should also adapt with learned {\em resource sensitivity} of different 
tasks which determine $\phi_j(\mathbf{x}, \mathbf{y})$. A key insight is that the running tasks can continuously provide resource sensitivity information, and the waiting tasks indicate increased completion time and future contention. Therefore, each edge cluster needs to constantly monitor resource utilization, running tasks, waiting tasks, and newly submitted tasks. To speed up the action searching, we define a set of pending tasks $\mathcal{P}_{n,t}$ with a fixed size 
as the input of our task selection phase. Each scheduler $n$ appends up to $|\mathcal{P}_{n,t}|$ tasks that have the earliest arrival times from the queue (i.e., $\in \mathcal{Q}_{n,t}$) to its pending set. 

By differentiating pending tasks and queuing tasks, we can reshape the decision space into one with a hierarchical structure, i.e., the optimizations of the pending set and running set are decomposed. The search space of selecting tasks from the queue to schedule is thus reduced, as the running tasks can only be chosen from the pending ones. Flexibility is also offered in that we can limit the size of the pending set in the edge cluster $n$ to a reasonable range (e.g., 10), to prevent excessive queue-cutting behavior.
First, we define the features of raw states for the agent $n$, namely $s_{n,t}=\{\mathcal{O}_{n,t}, \mathcal{P}_{n,t}, e_{n,t}, q_{n,t}\}$, which serves as the input of our HAN and is formalized as follows.
\begin{itemize}[leftmargin=*]
\item {\bf Running} task feature set $\mathcal{O}_{n,t}$ consists of entities $o_{j,n,t}$,
each of which is the concatenation of the resource allocation, task type, and elapsed time for a running task $j$ in $n$ and $t$. Since there are a total of $R$ types of resources and $J$ types of tasks, we have $o_{j,n,t} \in \mathbb{R}^{R+J+1}$, given that we use a one-hot format to indicate the task type. 
Note that $|\mathcal{O}^{n}_{t}|$ is the number of running tasks in the edge cluster in the timestep $t$, which varies over time. 
These running tasks not only provide real-time feedback on resource sensitivity but also indicate the future available resources that departing tasks will release. 
\item {\bf Pending} task feature set $\mathcal{P}_{n,t}$ contains entities $p_{j,n,t}\in \mathbb{R}^{R+J}$, each of which 
encodes the concatenation of the minimum resource requirement and the type of task $j$. We pad the pending set with dummy entities $p_0$, each of which has an extra binary digit representing whether the node is dummy or not. 

\item {\bf Resource} feature vector $e_{n,t} \in \mathbb{Z}^{R}$ 
represents the available resources of cluster $n$ in timestep $t$.
\item {\bf Queuing} task feature $q_{n,t}$ is a scalar of the number of remaining tasks in the queue excluding those in the pending set. It reflects the current system workload and thus has a strong correlation with the objective value.
\end{itemize}

In our MARL framework, if each agent $n$ can only observe the inner-cluster features $\{\mathcal{O}_{n,t}, \mathcal{P}_{n,t}, e_{n,t}, q_{n,t}\}$, they will be more prone to being trapped in their local optima. A global representation of features in all clusters is thus needed. However, we cannot simply stack $s_{n,t}$ over all $n$ into a global state matrix and feed it to each agent's model. The main drawbacks are: 1) the graph structure of the states will be lost; 2) the state space is not compact due to duplicate information such as the tasks shared in multiple $\mathcal{P}_{n,t}$ and $\mathcal{Q}_{n,t}$. Our solution instead embeds the entire graph into a neural network and enables iterative state interaction across clusters, with inter-cluster information reinforced over time based on importance.    

Since HAN is designed to embed heterogeneous nodes, link relations, and their semantics~\cite{wang2019heterogeneous}, it fits well with our state representations. 
As shown in Fig.~\ref{fig:system}, we design our HAN to be a graph $\mathcal{G}_t(\mathcal{V}_t, \mathcal{E}_t)$, where the node set $\mathcal{V}_t$ consists of $\mathcal{O}_t$, $\mathcal{P}_t$, cluster nodes $\mathcal{N}$, and a shortcut node. We abuse the notation of the task and cluster indices to denote the corresponding nodes in $\mathcal{G}_t(\mathcal{V}_t, \mathcal{E}_t)$ as well. The edges in $\mathcal{E}_t$ are defined as follows. A running task node $j\in \mathcal{O}_t$ connects with a cluster node $n\in \mathcal{N}$ if and only if task $j$ is running on cluster $n$. Analogously, a pending task node $j\in \mathcal{P}_{t}$ connects with the cluster node $n$ when task $j$ is in the pending set, i.e., $j\in \mathcal{P}_{n,t}$. Each $j$ can connect to multiple cluster nodes since they can be in the queues of multiple $n$, and we construct each $\mathcal{P}_{n,t}$ by dequeuing 
tasks in $\mathcal{Q}_{n,t}$ until $\mathcal{P}_{n,t}$ is full or $\mathcal{Q}_{n,t}$ is empty. The information propagation of our HAN follows the update steps in~\cite{wang2019heterogeneous} by passing the features as messages from the {\em neighbors} to each node $u\in \mathcal{V}_t$ and aggregating them with the features of $u$ using a two-level attention network in a configurable number of interactions. To speed up the message propagation between nodes that are far from each other, we add a shortcut node 
that connects all the cluster nodes to better adapt to a large number of agents. Similar to the virtual node in the GNN literature~\cite{gilmer2017neural}, the intuition is that adding a shortcut node can shorten the rounds of full message passing. The original connection of the cluster nodes and both types of task nodes form a bipartite graph. If the number of agents increases, the message may not be delivered to all other nodes under a certain number of rounds. So we use the shortcut node as a way to quickly pass long-range information.
The propagation model of our HAN is formalized below.
\begin{align}\label{eq:han_state}
\begin{split}
& G^{\mathrm{shortcut},(0)}_{t} = \bm{0} \\
& G^{N,(0)}_{t} = \cup_{n=1}^N \{(e_{n,t},q_{n,t})\} \\
& G^{\mathcal{P}, (0)}_{t} = \cup_{n=1}^N \mathcal{P}_{n,t} \\
& G^{\mathcal{O},(0)}_{t} = \cup_{n=1}^N \mathcal{O}_{n,t}
\end{split}
\end{align}
We concatenate $e_{n,t}$ and $q_{n,t}$ for all the clusters as the input features of the cluster nodes, and assign $p_{j,t}$ and $o_{j,t}$ to every pending and running task node respectively. We denote the initial global state input as $G^{(0)}_{t}=\{G^{\mathrm{shortcut},(0)}_{t}, G^{N,(0)}_{t}, G^{\mathcal{P}, (0)}_{t}, G^{\mathcal{O},(0)}_{t}\}$, as in~\eqref{eq:han_state}. The node embedding is propagated in each layer $l$, i.e., $G^{(l)}_{t} = g(G^{(l-1)}_{t})$, where $g(\cdot)$ represents the two-level attention network aggregating the features of each node with its neighbors. After $L$ layers of graph message passing, we have the final graph embedding $G^{(L-1)}_{t}$. We then map $G^{\mathcal{P}, (L-1)}_{t}$ to the corresponding agents as the input of their actor networks.

\subsection{Action Decomposition and Constraint Encoding} \label{subsec:decomposition}
Now we formalize our design of the NN architecture and functions for our actor network. The basic idea is to decompose the task placement and resource allocation decisions, since naively concatenating them together yields a much larger action space. Instead, we leverage their dependencies, i.e., we only need to allocate resources to the selected task (in \eqref{eq:st_x_y}). The challenges are using NN modules to encode such dependencies and enable them to learn our optimization constraints. We finally propose our training methods that can boost the agents to avoid conflicts and the decisions that violate constraints. 

\textbf{Pointer module.}
We design a pointer network inspired by the architecture of sequence-to-sequence NNs for natural language processing. The key insight is that we both need to solve problems in that the vocabulary of the output sequence will change with the length of the input sequence. Combinatorial problems like the traveling salesman problem usually have this trait~\cite{vinyals2015pointer, kool2018attention}. In our task selection phase, although we can fix the length of the input sequence with dummy entities, we cannot fix the vocabulary length of the output sequence in the same way, since the size of the pending set changes with time.
We need an extra mechanism, which enables dynamic vocabulary to avoid the undesirable and illegal situation of choosing a dummy entity. To this end, we implement a pointer network for decoding task selection actions by a functionally simplified attention layer. We formulate our pointer module for agent $n$ as follows. 
\begin{align}\label{eq:pointer}
& \bar{h}_{n,t} = \frac{1}{|\mathcal{P}_{n,t}|} \sum_{j: p_{j,n,t} \in \mathcal{P}_{n,t}}^{|\mathcal{P}_{n,t}|}{h_{j,n,t}} \notag\\
& \hat{u}_{j,n,t} =
  \begin{cases}
    -\infty, & \forall j: p_{j,n,t} = p_0 \\
    v^T \tanh(W_1 h_{j,n,t} + W_2 \bar{h}_{n,t}), & \text{otherwise.} \\
  \end{cases} \notag\\
& \hat{z}_{j,n,t} = \mathrm{softmax}(\hat{u}_{j,n,t})
\end{align}
Here, $h_{j,n,t}$ is the output of task $j$ of the transformer encoder. $W_1$, $W_2$, and $v^T$ are model parameters. We design attention scores of the model, $\hat{z}_{j,n,t}$, to represent the probability of scheduling a pending task $i$, shown as the output of the second \textit{Attn} in Fig.~\ref{fig:system}. We then sample the initial task selection action for the scheduler from the probability distribution constructed by $\hat{z}_{j,n,t}$, i.e., $z_{j,n,t}\sim \{\hat{z}_{j,n,t}\}_{j\in \mathcal{P}_{n,t}}$, e.g., the task labeled by \textit{a} is sampled by agent 1 in Fig.~\ref{fig:system}. We mask $\hat{u}_{j,n,t}$ of the dummy entities as $-\infty$ so that the corresponding $\hat{z}_{j,n,t}$ will be $0$ and thus the dummy entities will never be sampled. Since $z_{j,n,t}$ is the task selection action chosen by the agent $n$ before being coordinated with other agents, we have $z_{j,n,t}\ge y_{j,n,t}, \forall j,n,t$, where $y_{j,n,t}$ is defined as our final task placement decisions. 

\textbf{Task selection conflicts.}
Edge devices can submit offloading requests to multiple edge clusters and this is where conflicts may arise. We do not consider the scenario where each task is split onto and served by multiple edge clusters simultaneously, for the unstable network condition between the edge clusters. 
To resolve task selection conflicts, we transform the attention scores in the output of the task selection phase, into task-conditioned probabilities using Bayes' theorem. The attention scores $\hat{z}_{j,n,t}$ can be interpreted as the probability of selecting task $j$ in the task selection phase by the agent $n$, which means task $j$ was selected with a $\mathrm{Pr}(j|n)$ chance, which is the conditional probability of choosing the task $j$ given the agent $n$.
 For every agent that chooses task $j$, we use the maximum number of task $j$ that can be held in the cluster with the minimum required resources satisfied to approximate the marginal probabilities of choosing agent $n$, i.e., $\mathrm{Pr}(n)$.
The intuition is that considering load-balancing, we hope that the clusters with more available resources can more possibly get the conflicted task. Finally, the agent with the largest task-conditioned probability can schedule the task, and the agents that fail to schedule the task yield void actions in this scheduling interval.
\begin{align}\label{eq:bayes}
\begin{split}
& \mathrm{Pr}(j|n) = z_{j,n,t},\quad  \mathrm{Pr}(n|j) =  \frac{\mathrm{Pr}(j|n)\mathrm{Pr}(n)}{\sum_{n=1}^{N}\mathrm{Pr}(j|n)\mathrm{Pr}(n)} \\
& y_{j,n,t} = 
\begin{cases}
    1, & n = \argmax_{n'} {\mathrm{Pr}(n'|j)} \\
    0, & \text{otherwise.} \\
\end{cases}
\end{split}
\end{align}
Nonetheless, task selection conflicts bring non-negligible penalties to the algorithm in the {\em single-task scheduling} version. The reason is that, once multiple clusters choose the same task ($z_{j,n,t}=1$), only one of them can schedule the task in the current time interval $t$, and others will wait until $t+1$ to select new tasks due to their decision conflicts at $t$. This may result in an unnecessary waste of resources especially when the number of clusters is large, penalizing the total completion time. In Section \ref{subsec:multitask}, we manage to eliminate such penalties by extending the algorithm to {\em multi-task scheduling}, which can significantly improve the resource utilization of each edge cluster and thus the total task completion time.

\textbf{Invalid action masking.}
We do not use the reward penalty like \cite{tuli2020dynamic} to prevent the agents from taking invalid actions, because we observe an unstable training curve by giving a large negative reward to the agents predicting invalid actions, which is consistent with the findings of~\cite{huang2020masking}. Besides, adding penalty terms into rewards needs careful tuning of the associated parameters. Our intuition is that, the penalty should be large enough since invalid actions violate our constraints and thus need to be prohibited. But the value functions of different actions are approximated through an NN. Large penalties may also decrease the value estimates of other actions, especially those that are close to the invalid actions but turn out good or even optimal resource allocations.
To overcome this, we implement a different approach, which is to use an invalid action masking module to identify and mask all the invalid actions in the combinatorial action space to prevent our actors from predicting invalid resource allocation. We assign 0 probability to all the actions that either exceed the resource capacity or are less than the minimum required amount of resources.

\textbf{Loss function and training method.}
In our algorithm, the granularity of the scheduling intervals is relatively small compared to the long-running ML tasks. 
As a result, in some timesteps, the edge clusters may either be idle or lack resources to schedule. We skip all those timesteps for training since they contain little information for the agents to learn and bring noise and instability to the training. 
We use the multi-agent proximal policy optimization (MAPPO)~\cite{schulman2017ppo} to train our MARL model. The main reason why we use PPO is that we prefer a stable on-policy learning algorithm that can easily handle the large combinatorial action space and is sufficiently general to adapt to volatile network environments. To design our training loss function, we adopt the clipped surrogate loss function of PPO as the base, and we 
make several modifications to the base loss function in order to 
customize it for our optimization problem. First, we mask all the invalid resource allocations, i.e., decisions $x_{j,n}$ that exceed the resources capacities $C_{r,n}$ given the minimum resource requirements $b_{j,r}$ of controlled tasks. The masked probabilities are defined in \eqref{eq:loss}. Then the scheduling policy can be derived from the chain rule of masked conditional probabilities, formulated in \eqref{eq:loss_chain}. We further use $m$ as an indicator to mask the final resource allocation predictions $\mathrm{Pr}^*(x_{j,n}|j,n)$ of the conflicted agents that fail to schedule the task, while keeping the task selection predictions $\mathrm{Pr}(j|n)$ in the loss function.
\begin{align}\label{eq:loss}
& \mathrm{Pr}^*(x_{j,n}|j,n) = \mathrm{MASK}(\mathrm{Pr}(x_{j,n}|j,n), b_{j,r}, C_{r,n}) \\
& \pi_{\theta_n}(\bm{a}_n|\bm{s}_n) = \mathrm{Pr}(j|n) \mathrm{Pr}^*(x_{j,n}|j,n)^{m} \label{eq:loss_chain}\\
& m = 
\begin{cases}
    1, & y_{j,n,t} = 1, \\
    0, & \text{otherwise.} \\
\end{cases}
\end{align}

We train our MARL model offline because online training is time-consuming and may easily lead to poor performance if not well tuned~\cite{peng2021dl2}. For every policy update step, we collect a batch of trajectories $(\bm{s}, \bm{a}, \bm{r})$ from the environment,
each representing the concatenation of the states, the actions and the rewards for all the agents. Every agent learns a policy $\pi_{\theta_n}(\bm{a}_n|\bm{s}_n)$, which is a joint distribution of the combinatorial actions giving the states.

\subsection{Multi-task Scheduling} \label{subsec:multitask}
Like {\em TapFinger}, some of the current works schedule only one task at a real-world timestep onto each cluster~\cite{zhao2021large, han2021kais}, considering that a short interval between the timesteps can overlap with the scheduling overhead. This was referred to as {\em single-task scheduling} in the previous sections. However, scheduling multiple tasks simultaneously to the clusters at the beginning of each real-world interval can achieve higher system throughput and thus better task completion time. Compared with {\em single-task scheduling}, two main advantages are summarized below. 

{\em More opportunities to resolve conflicts.} As elaborated in Section \ref{subsec:decomposition}, the clusters do not have any new tasks to serve in a real-world interval once being conflicted at the corresponding timestep. But if we could allow each cluster to choose multiple tasks simultaneously, it would potentially have more chances to choose tasks that avoid conflicts. 

{\em Better resource allocation by packing tasks.} Even if there is no conflict at some timestep, clusters with sufficient resources can support more than one task. The resource utilization will potentially be improved, if the agents allocate the prevailing resources to multiple tasks simultaneously. Note that the optimal scheduling for total completion time minimization may not use up the currently available resources for scheduling a single ML task. 

{\bf Design intuition.} To maintain a relatively small action space, we proposed to choose tasks one by one sequentially. Our design is then driven by the intuition that: in the previous {\em single-task scheduling}, each timestep $t$ corresponds to a real-world time, which causes a waste of resources on conflicted clusters, since they have no new tasks to serve for one real time slot. But we could instead divide a real-world time interval into multiple logical timesteps, each of which represents the step of making a decision for selecting one task. If a cluster loses the chosen task due to conflicts, it continues to the next logical timestep with negligible overhead, e.g., about 150ms (see Section~\ref{sec:experiment}), until getting one task to run in the current interval or having an empty queue. In this way, our proposed MARL framework and NN design can be naturally extended to serve multiple tasks simultaneously in each real-world time for each cluster. 
Agents can do multiple inferences to schedule tasks to their greatest extent and hence fully utilize the available resources in the clusters.

\begin{figure}[t]
  \includegraphics[width=0.8\linewidth]{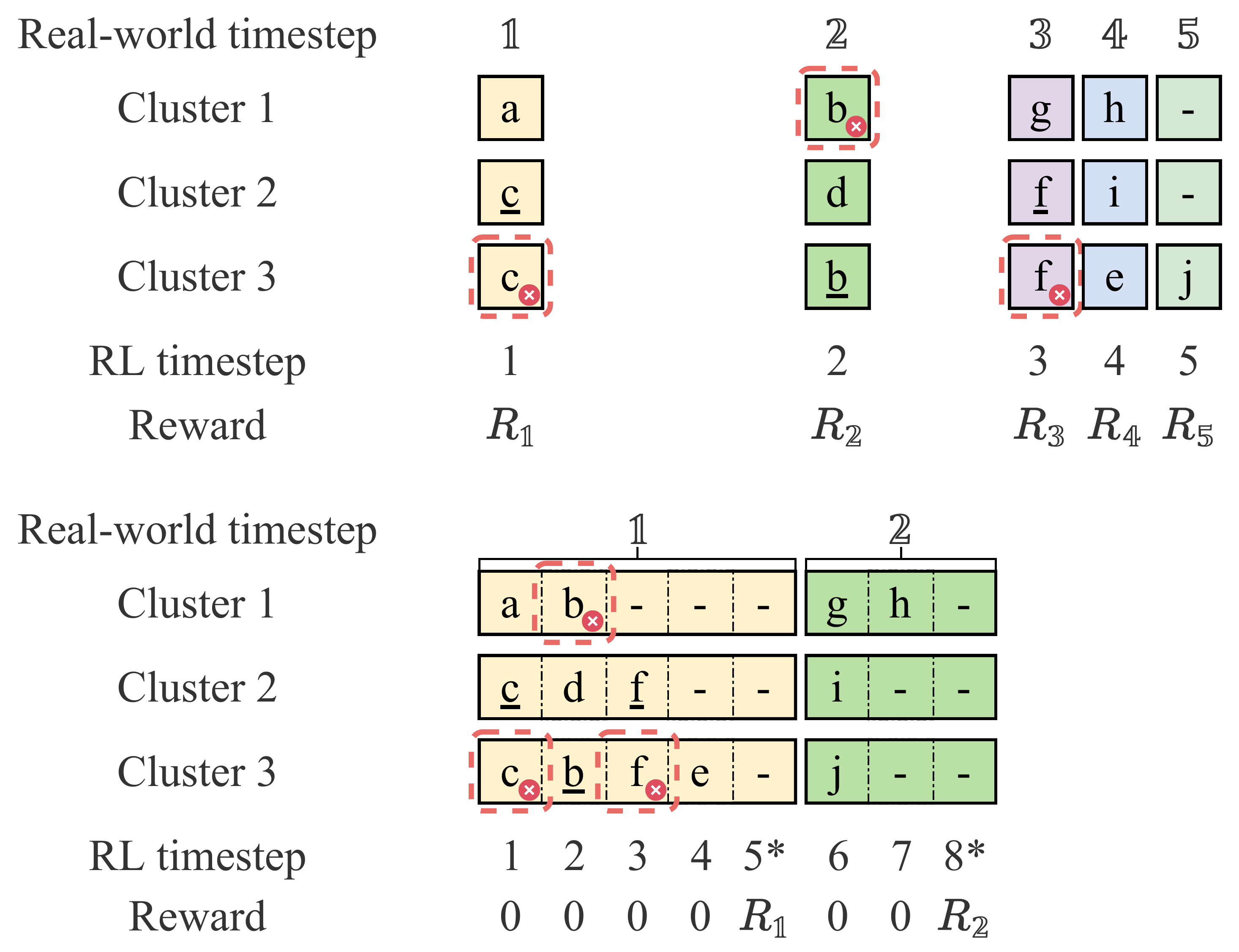}
  \centering
  \caption{Differences in the scheduling process between the single-task {\em TapFinger} (top) and the multi-task {\em TapFinger} (bottom) in a contrived example.}
  \label{fig:multi-task}
\end{figure}

{\bf Illustrative example.} We begin to illustrate the multi-task extension with a contrived example. We consider 3 edge clusters and a workload in which all the tasks can be accomplished in 1 real-world time interval. The edge clusters are capable of simultaneously running multiple tasks if their resources are sufficient. We use single-task {\em TapFinger} and multi-task {\em TapFinger} to refer to our implementation for the {\em single-task scenario} and {\em multi-task} scenario, respectively.
Fig.~\ref{fig:multi-task} shows the scheduling process of the single-task {\em TapFinger} (top) and the multi-task {\em TapFinger} (bottom) using the same workload trace. We use lowercase letters to denote the tasks and a hyphen to denote a void action. In the single-task {\em TapFinger}, each edge cluster schedules one task at a time. 
However, some of the clusters may have sufficient resources to support multiple tasks simultaneously in a single real-world time interval. Moreover, the clusters that encounter a conflict and lose their scheduling opportunities (marked with red dotted boxes and diagonal crosses) will pass a scheduling interval using the single-task {\em TapFinger}, leading to sub-optimal overall task performance.

{\bf Algorithm interpretation.} Our multi-task {\em TapFinger} can schedule multiple timesteps in a real-world time interval, and schedule other tasks when conflicts occur. To realize this without expanding the action space, we keep the design of scheduling only one task in one RL timestep. As presented in Algorithm \ref{alg:multi_task}, after taking scheduling actions, all the agents in the multi-task {\em TapFinger} update the states of the edge clusters as the input of the next RL timestep but hold the execution of the scheduled tasks until the next real-world interval. The agents keep scheduling tasks while they can only forward the execution at a global agreement where all of the agents yield void actions. Intuitively, they may schedule more than one non-conflicted task in the same interval, based on the current resource availability and learned resource sensitivities of different ML tasks. In each RL timestep, the agents observe the actual rewards at the global agreement but 0 rewards for the scheduling actions. In our evaluation, a discount factor less than 1 will introduce biases among the scheduling actions made in the same real-world timestep, and slow down the convergence during training. Since these actions in the multi-task {\em TapFinger} are of equal importance, we set the discount factor to 1.

\begin{algorithm}[t]
 \caption{The multi-task {\em TapFinger}}\label{alg:multi_task}
 \begin{algorithmic}[1]
 \STATE \textbf{Input}: initial state $\mathbf{s}$
 \STATE $\mathrm{continue}\leftarrow \mathrm{true}$
 \STATE For each agent $n \in [N]$, $x_{n}\leftarrow \mathbf{0}$, $y_{n} \leftarrow \mathrm{NULL}$
 \WHILE{$\mathrm{continue} = \mathrm{true}$}
    \STATE $\mathrm{continue} \leftarrow \mathrm{false}$
    \STATE For each agent $n \in [N]$, $(x_{n}, y_{n}) \leftarrow \mathrm{TapFinger}(\mathbf{s},n)$
    \STATE $\mathbf{x},\mathbf{y} \leftarrow \mathrm{ConflictResolution}(\mathbf{x},\mathbf{y})$
    \STATE $\mathbf{s} \leftarrow \mathrm{Apply}(\mathbf{x},\mathbf{y})$
    \FOR{agent $n \in [N]$}
        \IF{$y_{n} \neq \mathrm{NULL}$}
            \STATE $\mathrm{continue} \leftarrow \mathrm{true}$
        \ENDIF
    \ENDFOR
 \ENDWHILE
 \STATE The execution of the scheduled tasks starts
\end{algorithmic} 
\end{algorithm}

As shown in Fig.~\ref{fig:multi-task}, the multi-task {\em TapFinger} can achieve better performance in the case where the task length is small and the resources are sufficient. However, the DRL model of the multi-task {\em TapFinger} may require a longer trajectory and thus longer training time, e.g., some of the agents have already made their scheduling decisions but have to wait until the global agreement to forward the execution, i.e., the start of the next real-world time interval. In essence, we are trading the global trajectory length with real-world efficiency. Similar approaches are adopted in \cite{peng2021dl2,mao2016resource} for the single agent DRL. We find that this approach is also effective in multi-agent DRL as long as we only allow forwarding the execution when all the agents agree to it.
Such an extension of {\em TapFinger} has two advantages. First, by allowing multi-task scheduling, the edge clusters that support scheduling multiple tasks simultaneously can benefit from it and achieve larger system throughput. Second, the stalling effect caused by the conflicts among the agents can be amortized within a real-world timestep.

\section{Evaluation}\label{sec:experiment}
We evaluate both single-task {\em TapFinger} and multi-task {\em TapFinger} using synthetic and test-bed ML tasks. Both versions of {\em TapFinger} achieve considerable average completion time reduction in the corresponding environments. Experimental results also demonstrate the scalability of {\em TapFinger}, as it outperforms baselines with increasing network scale and system workload.

\subsection{Experiment Settings}
\textbf{Model configurations.}
We implement {\em TapFinger} with PyTorch and use Tianshou~\cite{tianshou}, an RL library based on PyTorch, to manage the training process. Our simulation environment implements the Gym~\cite{gym} standard interface to communicate with MARL agents. For the HAN implementation, we use PyTorch Geometric library~\cite{fey2019pyg} to accelerate the data loading, training, and inference for HAN. 
We use 6 HAN layers with 4 heads for multi-head attention to generate the HAN embedding with a hidden size of 256. By default, we set the size of the pending task set as 10. The pointer network module has a 2-layer transformer encoder with 4 heads for multi-head attention, and a functionally simplified attention layer to predict task selection actions. We use a 2-layer perceptron for resource allocation actions. As for the critic, we use a 2-layer perceptron that accepts the flattened HAN embedding of all the pending tasks in the system as the input. A server with $1\times$ Intel i9-12900K CPU and $1\times$ NVIDIA RTX 3080 GPU is used to train our MARL model with 32 parallel training environments for either synthetic or test-bed ML tasks.

\textbf{Baselines.}
We regard CPU cores and GPUs as critical computation resources for ML training and inference tasks, and we assume that the edge clusters only allocate the requested amounts of memory to the tasks, since we observe little performance gain beyond the memory requirement of the tasks. The training tasks with multiple allocated GPUs are running in a distributed data-parallel fashion via all-reduce, so the schedulers need not specify parameter servers for them.
We implement two representative ML task scheduling algorithms and two heuristics as our baselines. 
We add a switching overhead for all the scheduling algorithms enabling dynamic resource allocation, to simulate the process of saving checkpoints, reallocating resources and resuming training or inference for ML tasks. Although the switching overheads are different for tasks using different amount of computing unit, we fix the switching overhead to 0.5 timestep for simplicity.
But even if we ignore the switching overhead, multi-task {\em TapFinger} can still outperform Tiresias and maintain comparable performance as Optimus, as shown in \ref{subsec:multi-task_experiment}.
\begin{itemize}[leftmargin=*]
\item Optimus. It uses curve-fitting performance models to estimate training speed as a function of the number of parameter servers, the number of workers, and batch size in each task. 
At the beginning of the scheduling process, it allocates 1 parameter server and 1 worker for each task. Then it incrementally chooses the allocation decision with the largest marginal gain for each task, until all the resource is used up or the marginal gains of all the tasks become non-positive.
To tailor Optimus to our problem, we consider CPU and GPU as the two types of resources and fix the batch size variable in the function to a constant. Then we use the same way as theirs to fit the progress function directly from the environments and generate the scheduling decisions.
\item Tiresias. It is a preemptive scheduling algorithm based on the least-attained service and multi-level feedback queue. It divides the queue according to the GPU time thresholds and has the tasks of each queue sorted in a First-In First-Out order of their start times. In our implementation, we use a 2-level feedback queue as the original paper recommended and the median GPU time as the threshold. 
\item Random/Minimum (Min) allocation. They randomly choose a task from the pending set and allocate a random/minimum but valid amount of resource to the chosen tasks.
\end{itemize}

\textbf{Synthetic ML tasks.}
To capture the heterogeneous and non-stationary resource sensitivity of the ML tasks, we define two types of synthetic ML tasks that vary greatly in duration time and resource sensitivity. We fix the progress gain of the resource-insensitive tasks for every resource allocation. The resource-insensitive tasks can be accomplished in 4 timesteps as long as the minimum resource requirement is met. Conversely, the resource-sensitive tasks can be accomplished from 2 to 17 timesteps under different resource allocations. The speed of the resource-sensitive tasks grows sub-linearly with the number of GPUs. We use the hyperbolic tangent function as the speed function, which shares some functional properties with the formulation of Amdahl's law~\cite{amdahl1967validity}. 

\begin{table}[!htb]
\centering
\caption{Test-bed ML tasks.}
\resizebox{\linewidth}{!}{%
\begin{tabular}{|c|c|c|c|c|}
\hline
ML phase  & Task                 & Model       & Batch size             & Duration \\ \hline
Training  & Image classification & ConvNet     & 1024                   & 34-262s  \\ \hline
Training  & Language modeling    & Transformer & 16384 (tokens) per GPU & 75-264s  \\ \hline
Inference & Speech recognition   & Wav2Vec2    & 1                      & 17-18s   \\ \hline
\end{tabular}%
}
\label{table:tasks}
\end{table}

\textbf{Test-bed ML task traces.}
We execute the test-bed ML training and inference tasks in a server with $4\times$ Intel Xeon Gold 6348 CPU and $8\times$ NVIDIA RTX 3090 GPU. We implement distributed data parallelism for a total of 3 ML training and inference tasks with PyTorch, as shown in Table \ref{table:tasks}.
To our knowledge, the retraining and inference tasks on the edge typically take up seconds to a few minutes\cite{Ekya2021NSDI}. Therefore, the task durations we set for both synthetic and test-bed ML tasks are relatively shorter than in the existing literature. Nonetheless, scheduling a large number of short-duration tasks is more delicate and offers more optimization opportunities. Besides, our proposed algorithm can adapt to longer and more diverse task duration and trajectories effortlessly.
We run the tasks in an ML task workload with every resource allocation combination, ranging from the minimum resource requirement of 2 CPU cores and 1 GPU, to 16 CPU cores and 8 GPUs. We use the PyTorch interface to control the number of GPU devices for the tasks. To control the CPU allocation with low overhead, we set the thread affinity policy according to the CPU allocation to constrain the OpenMP threads and the data loader threads on the designated CPU cores.
We downsize the running time of the tasks in the trace data similarly as \cite{bao2019deep}. The tasks are accomplished when the validation loss reaches the preset target for the training tasks or when the outputs of all the samples are calculated for the inference tasks. We collect the validation loss of the tasks and the elapsed time for every epoch and later use these data to shape the progress functions of the tasks over time. 

\textbf{Task arrivals.}
Our MARL models are trained with finite-size ML task workloads. We set the number of tasks in the ML workloads to be 64 times the number of agents in the environment. The simulated workloads follow a Poisson distribution with an arrival rate $\lambda=2$. 
We assign a bandwidth uniformly at random chosen from $[0,10]$ Mbps to each connection between the edge devices and the edge clusters, according to the measurements in~\cite{xu2021imc}. The edge device can be considered connectable to an edge cluster only if the bandwidth and the latency of their connection meet the minimum requirement for offloading, e.g., 5 Mbps.

\subsection{Performance of the Single-task {\em TapFinger}}
We train our MARL models for 8 million steps and save the models that achieve the best evaluation results. The scheduling interval (the real-world time interval between two successive executions of the MARL actions) is set to 10 seconds. In fact, this scheduling interval can be minimized according to the network transmission condition to achieve more timely task scheduling. Our measurements verify that the overhead of scheduling one task is about 0.3 seconds (elaborated in Section \ref{subsubsec:testbed}). Therefore, {\em TapFinger} can in fact achieve a scheduling interval of less than 1 second instead of 10 seconds, at the cost of a larger proportion (Fig.~\ref{fig:overhead_ratio}) of the overhead for running the scheduling algorithm.

We add $\pm0.1$ randomness to the normalized progress gains to simulate the unstable network condition between the edge devices and the edge clusters. The minimum resource requirements of both synthetic and test-bed ML tasks are 2 CPU cores and 1 GPU. Given that Fig.s~4 and 5 in \cite{Tiresias2019NSDI} demonstrate at least 5 seconds in pausing and resuming a data-parallel ML task, we set the overhead of adjusting the resource allocation for our baselines to be 5 seconds, which is equal to 0.5 timesteps. 

\begin{figure*}[!t]
    \begin{subfigure}{0.25\linewidth}
    \includegraphics[width=\linewidth]{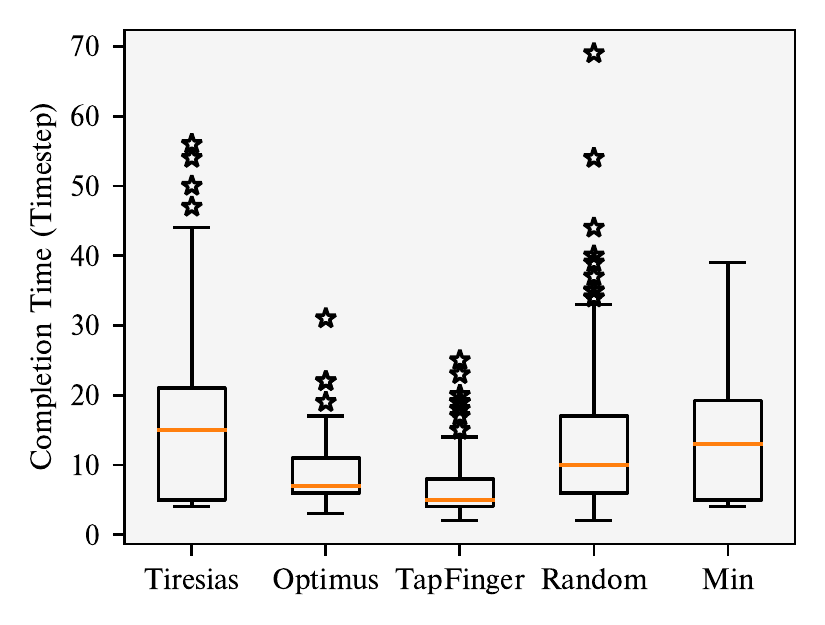}
    \caption{3 agents, 192 tasks.}
    \label{fig:synthetic_baseline_jct}
    \end{subfigure}\hfill
    \begin{subfigure}{0.25\linewidth}
    \includegraphics[width=\linewidth]{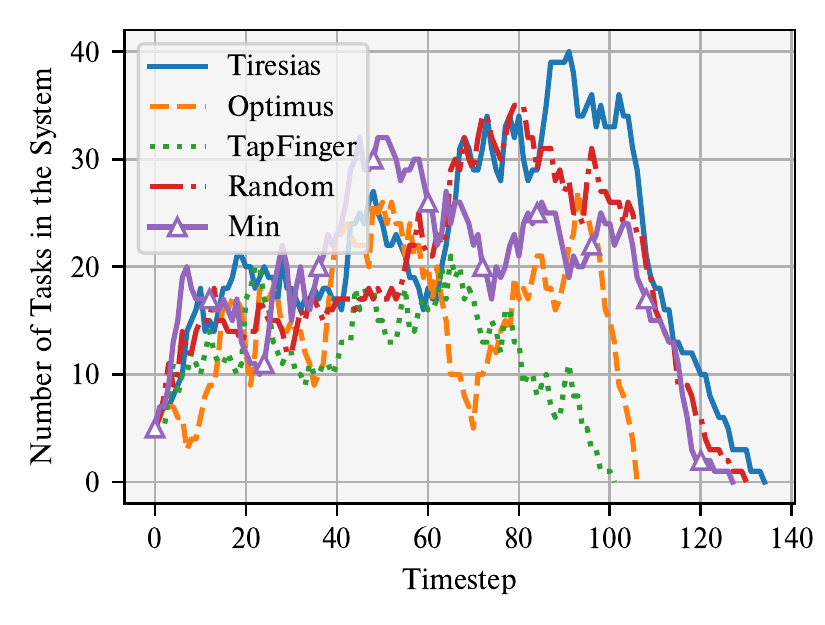}
    \caption{3 agents, 192 tasks.}
    \label{fig:synthetic_baseline_ta}
    \end{subfigure}\hfill
    \begin{subfigure}{0.25\linewidth}
    \includegraphics[width=\linewidth]{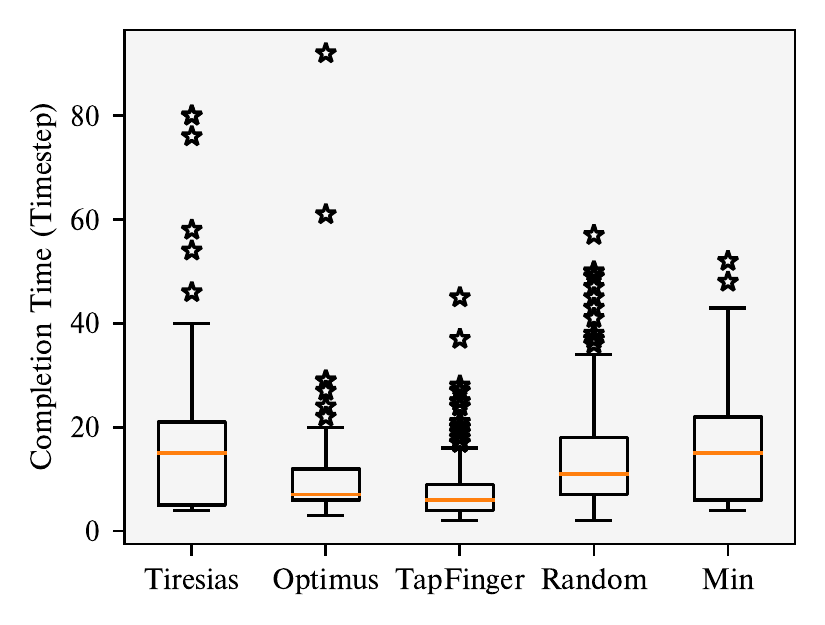}
    \caption{6 agents, 384 tasks.}
    \label{fig:synthetic_6_agent_jct}
    \end{subfigure}\hfill
    \begin{subfigure}{0.25\linewidth}
    \includegraphics[width=\linewidth]{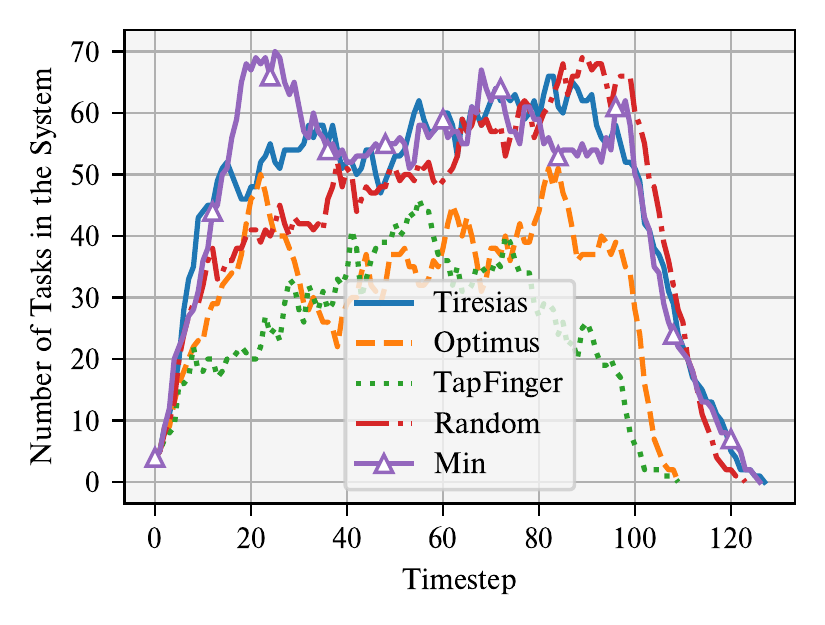}
    \caption{6 agents, 384 tasks.}
    \label{fig:synthetic_6_agent_ta}
    \end{subfigure}\hfill
    \caption{Single-task {\em TapFinger}: completion time and task accumulation comparison on synthetic ML workloads.}
    \label{fig:synthetic_baseline}
\end{figure*}

\subsubsection{Evaluation on Synthetic Data}\label{subsubsec:synthetic}
We first evaluate the single-task {\em TapFinger} on the synthetic ML tasks. We assume that each edge cluster has 16 CPU cores and 16 GPUs.

{\bf Three-agent implementation.} Fig.s~\ref{fig:synthetic_baseline_jct} and \ref{fig:synthetic_baseline_ta} show the completion time and the task accumulation of {\em TapFinger} as well as the baselines in a 3-agent environment. 
Task accumulation is the number of tasks that have arrived in the system at each timestep.
Because of the pre-run estimation of the progress function, Optimus performs better than other baselines.
Tiresias assumes unknown task characteristics and only schedules tasks given the minimum resource requirements, resulting in inferior performance. 
We observe a $28.6\%$ reduction in the median completion time and $25.1\%$ reduction in the average completion time of {\em TapFinger}, in comparison with Optimus.
Fig.~\ref{fig:synthetic_baseline_ta} also shows that {\em TapFinger} achieves the least task accumulation. 


\begin{figure}[!t]
    \centering
    \minipage[t]{0.5\linewidth}
    \includegraphics[width=\linewidth]{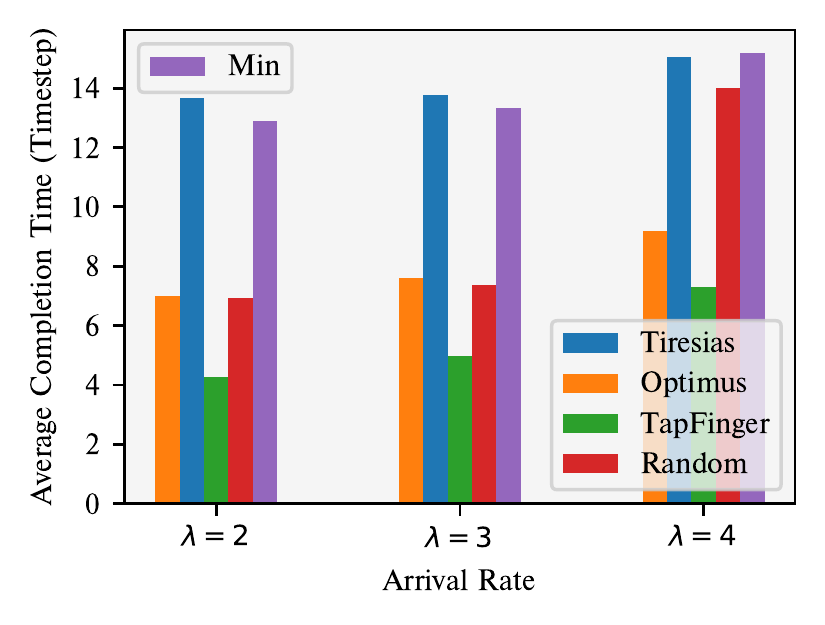}
    \caption{Varying arrival rates.}
    \label{fig:avg_jct_ar}
    \endminipage\hfill
    \minipage[t]{0.5\linewidth}
    \includegraphics[width=\linewidth]{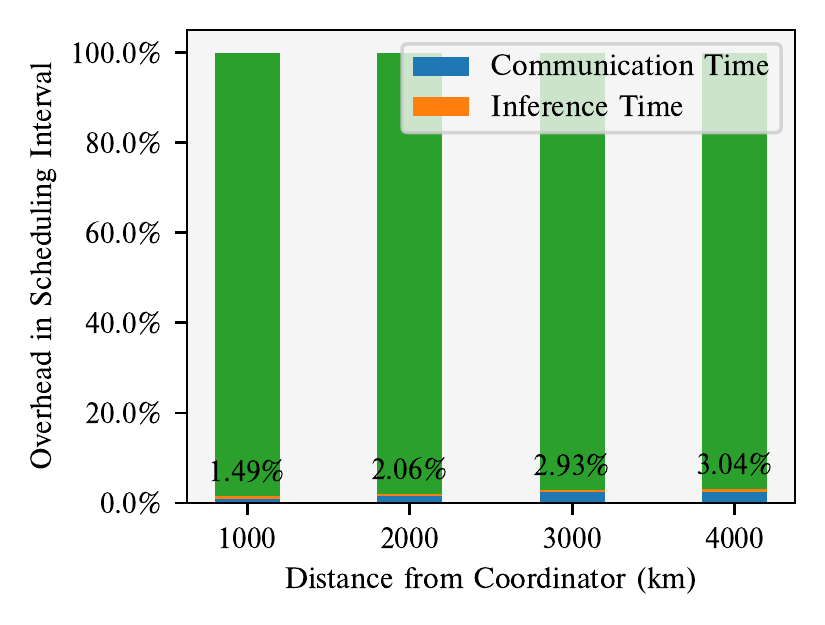}
    \caption{Scheduling overhead.}
    \label{fig:overhead_ratio}
    \endminipage\hfill
\end{figure}

{\bf Six-agent implementation.} We scale {\em TapFinger} to an environment with 6 edge clusters and an arrival rate of 4. As shown in Fig.s~\ref{fig:synthetic_6_agent_jct} and \ref{fig:synthetic_6_agent_ta}, {\em TapFinger} still shows at least 20.5\% improvement on the completion time and much lower task accumulation, compared with the baselines. We then test the robustness to the task arrival rate, ranging from 4 to 2, while keeping other settings unchanged. Fig.~\ref{fig:avg_jct_ar} shows that the task accumulation of {\em TapFinger} also outperforms other algorithms significantly, indicating that {\em TapFinger} is robust to varied volumes of the workloads once trained with sufficient workloads. 
We also observe a performance degradation of Optimus and Tiresias in these less heavy workloads. It can be explained by the fact that frequently reassigning resource allocation for each task leads to severe overhead due to preemptions.

\begin{figure}[!t]
    \centering
    \begin{subfigure}{0.5\linewidth}
    \includegraphics[width=\linewidth]{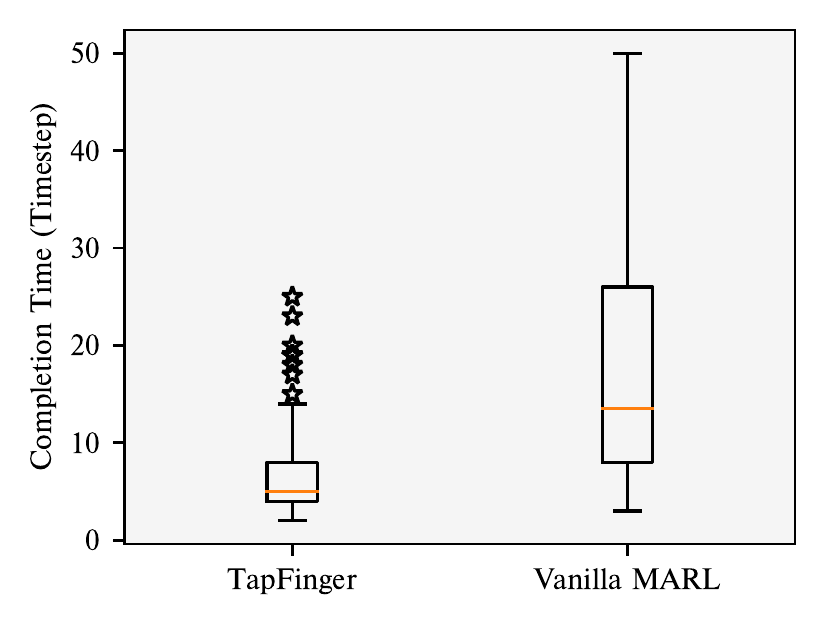}
    \caption{Completion time.}
    \label{fig:synthetic_toy_jct}
    \end{subfigure}\hfill
    \begin{subfigure}{0.5\linewidth}
    \includegraphics[width=\linewidth]{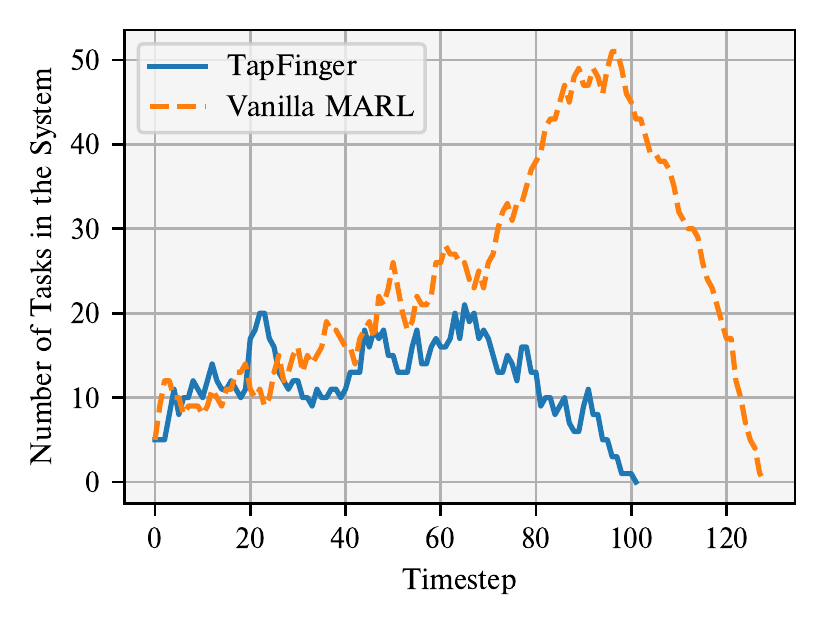}
    \caption{Task accumulation.}
    \label{fig:synthetic_toy_ta}
    \end{subfigure}\hfill
    \caption{Compared with the vanilla MARL on synthetic data.}
    \label{fig:synthetic_toy}
\end{figure}

{\bf Comparing with vanilla MARL.} We compare {\em TapFinger} with a vanilla MARL algorithm without HAN and the pointer network, which uses stacking state features and chooses the task from the queue in a First-In First-Out manner and predicts the resource allocation with our invalid action masking design. As implied in Fig.~\ref{fig:synthetic_toy}, the vanilla MARL algorithm struggles to recognize the task characteristics and avoid task selection conflicts.


\begin{figure*}[!t]
    \begin{subfigure}{0.25\linewidth}
    \includegraphics[width=\linewidth]{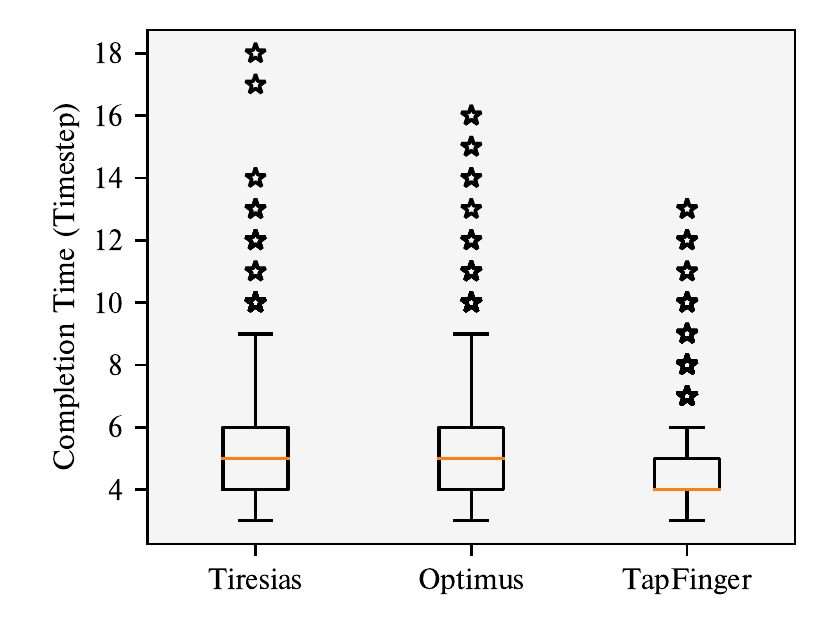}
    \caption{3 agents, 1000 tasks.}
    \label{fig:trace_3_agent_jct}
    \end{subfigure}\hfill
    \begin{subfigure}{0.25\linewidth}
    \includegraphics[width=\linewidth]{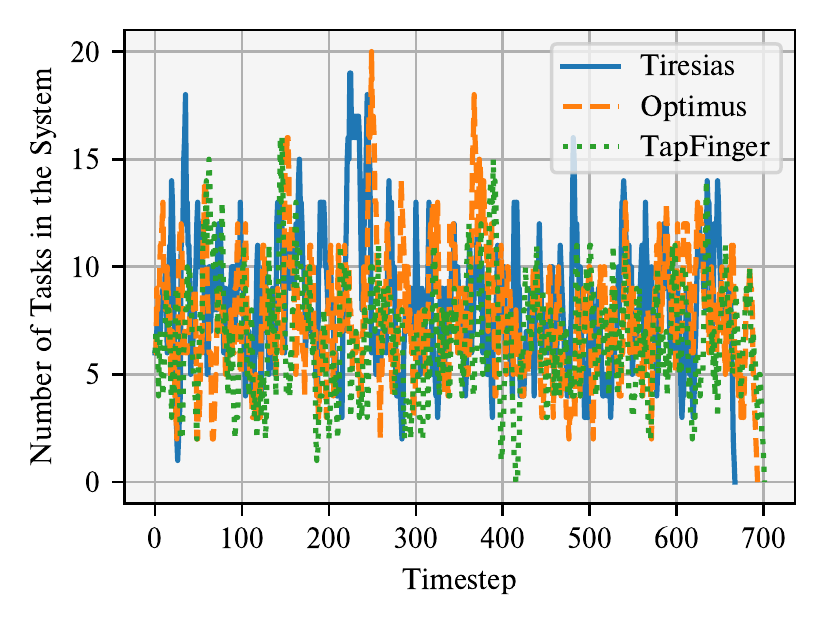}
    \caption{3 agents, 1000 tasks.}
    \label{fig:trace_3_agent_ta}
    \end{subfigure}\hfill
    \begin{subfigure}{0.25\linewidth}
    \includegraphics[width=\linewidth]{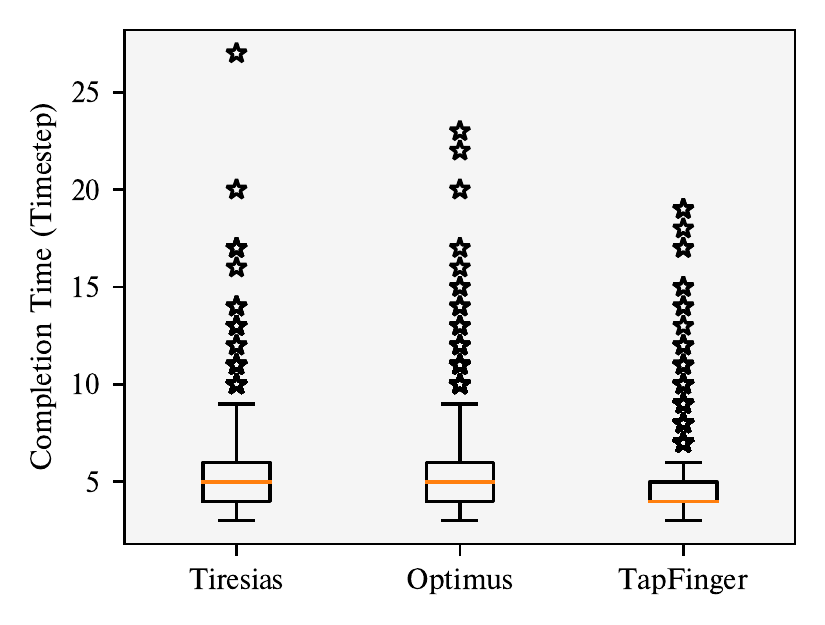}
    \caption{6 agents, 2000 tasks.}
    \label{fig:trace_6_agent_jct}
    \end{subfigure}\hfill
    \begin{subfigure}{0.25\linewidth}
    \includegraphics[width=\linewidth]{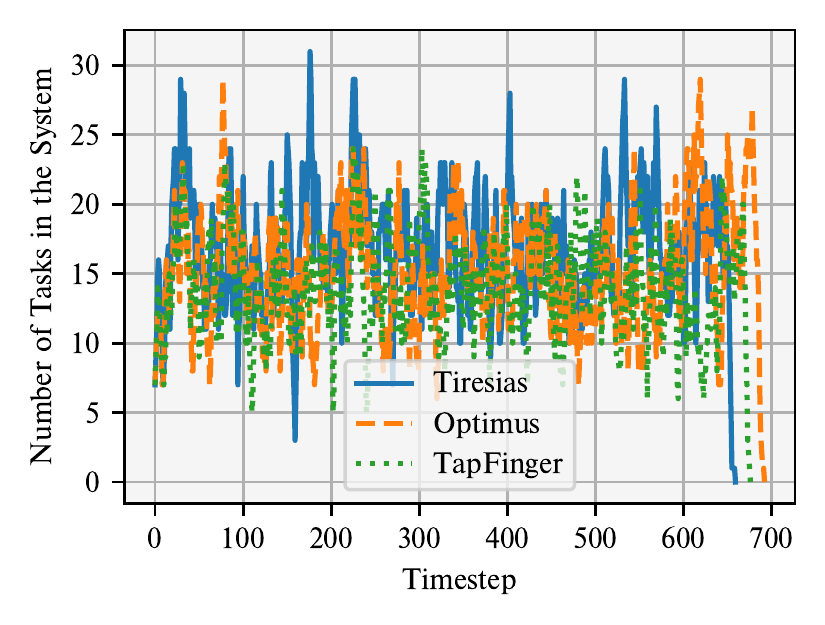}
    \caption{6 agents, 2000 tasks.}
    \label{fig:trace_6_agent_ta}
    \end{subfigure}\hfill
    \caption{Single-task {\em TapFinger}: completion time and task accumulation comparison on test-bed ML workloads.}
    \label{fig:testbed_baseline}
\end{figure*}

\begin{figure}[!t]
    \centering
    \begin{subfigure}{0.5\linewidth}
    \includegraphics[width=\linewidth]{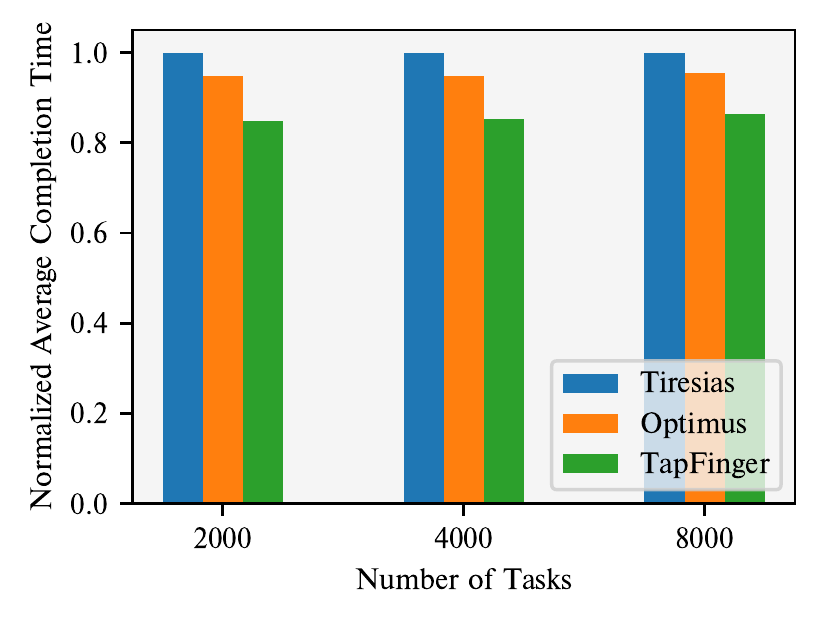}
    \caption{Increasing workload scale.}
    \label{fig:avg_jct}
    \end{subfigure}\hfill
    \begin{subfigure}{0.5\linewidth}
    \includegraphics[width=\linewidth]{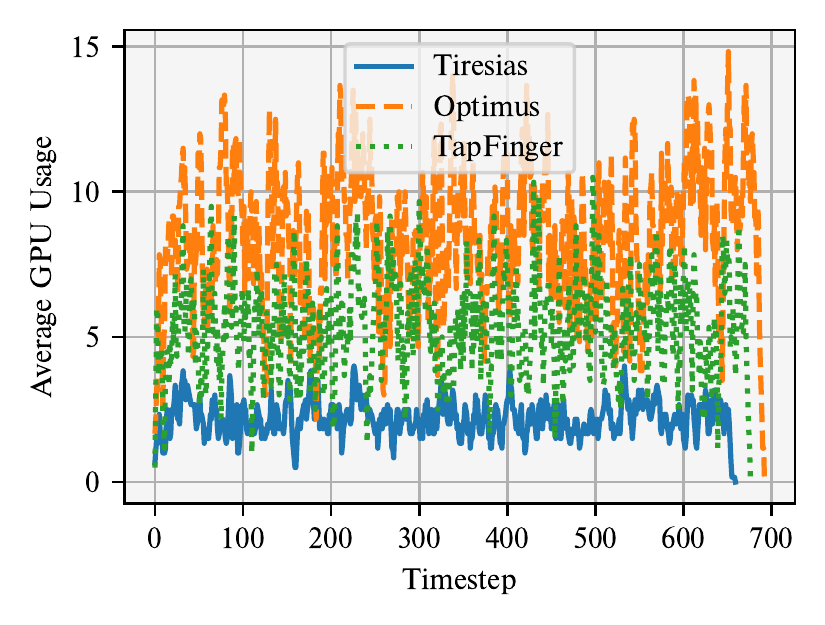}
    \caption{GPU utility.}
    \label{fig:gpu_usage}
    \end{subfigure}\hfill
    \caption{Single-task {\em TapFinger} in long-running workloads.}
    \label{fig:long_running_trace_jct}
\end{figure}

\subsubsection{Evaluation on Test-bed Data}
\label{subsubsec:testbed}
We further conduct several long-running experiments on test-bed ML workloads and evaluate the overhead incurred by running the MARL algorithm. 
The 3-agent and 6-agent cases of {\em TapFinger} are trained with an arrival rate of 4 and 6, respectively. Each edge cluster has 16 CPU cores and 8 GPUs.
We evaluate the algorithms in longer workloads than that in the training stage. Accordingly, we raise the resource capacity of each edge cluster to 32 CPU cores and 16 GPUs per agent. The arrival rates of the 3-agent and 6-agent test-bed environments are 1.5 and 3, respectively. 

{\bf Scheduling overhead.} We have measured that the total data size that needs to be transmitted in each real-world time interval as follows. The state observation and the action information take up 2KBs and 800Bytes respectively for each agent. We simulate the communication overhead using NetEm Linux kernel module~\cite{hemminger2005netem}. We set the bandwidth between the coordinator and the edge clusters as 1Mbps, and we vary the network latency by increasing geographical distances according to the measurements of Alibaba Cloud edge clusters \cite{xu2021imc}. Fig.~\ref{fig:overhead_ratio} shows that the inference and communication time of {\em TapFinger} is negligible compared to the scheduling interval. 

Fig.~\ref{fig:testbed_baseline} shows that the single-task {\em TapFinger} is scalable with the increasing number of agents, and it achieves a considerable average completion time reduction of 13.7\% and 10.4\% over the baselines in the corresponding environment settings.
It is also applicable to various lengths of workloads. As shown in Fig.~\ref{fig:avg_jct}, the 6-agent {\em TapFinger} maintains 14.5\% and 10\% reduction on the average completion time compared with Tiresias and Optimus, even with a longer workload scale than that in the experiments for Fig.~\ref{fig:testbed_baseline}. Considering that the test workloads are all longer than that for our offline training, the results indicate that a well-trained {\em TapFinger} can adapt to online unseen and volatile workloads for practical edge cluster scheduling scenarios.
We also examine the GPU usage of the 6-agent {\em TapFinger} in the 2000-task workload. Fig.~\ref{fig:gpu_usage} shows that {\em TapFinger} strikes a good balance between resource efficiency and the completion time performance, since it achieves the best performance with moderate amounts of resources.

\subsection{Multi-task Scheduling and Heterogeneous Clusters}\label{subsec:multi-task_experiment}
We evaluate the multi-task {\em TapFinger} based on the same 3-agent and 6-agent synthetic workload settings as in Section~\ref{subsubsec:synthetic}. 

{\bf Scheduling overhead.} As explained for Fig.~\ref{fig:overhead_ratio} in Section \ref{subsubsec:testbed}, the time of a single inference using {\em TapFinger} is less than $3.04\%$ of the communication time (round-trip time) between geographical edge clusters. Hence the overhead of the serial inferences for scheduling multiple tasks simultaneously using the multi-task {\em TapFinger} does not increase much, given the same communication time as in the single-task scenario. Specifically, there is only one state information transmission between the coordinator and the edge clusters at the beginning of a scheduling (real-world) interval. The serial inferences can be run independently in the coordinator since the state transition within the real-world time interval can be inferred and generated from the predictions (i.e., actions) of the last logical RL timestep. 

\begin{figure}[!t]
    \centering
    \begin{subfigure}{0.5\linewidth}
    \includegraphics[width=\linewidth]{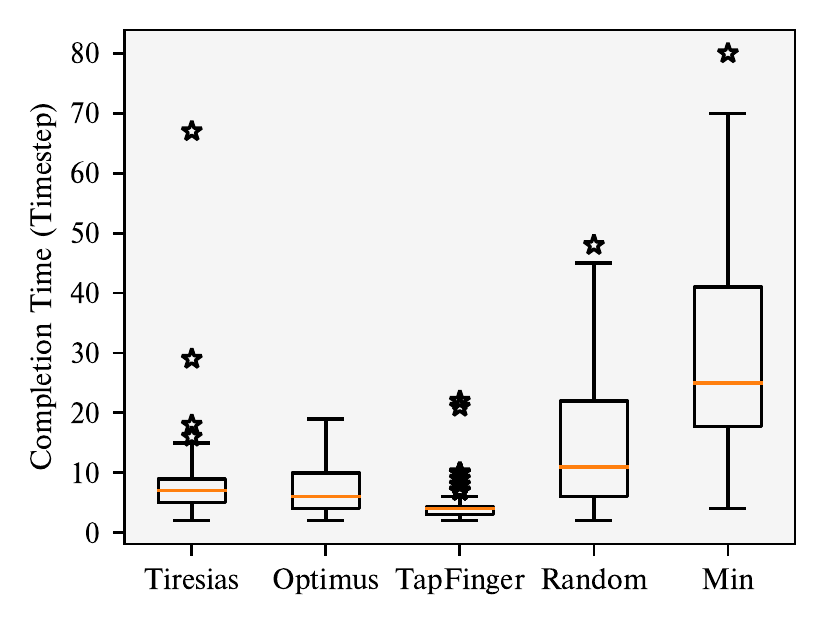}
    \caption{Completion time.}
    \label{fig:multiple_jct}
    \end{subfigure}\hfill
    \begin{subfigure}{0.5\linewidth}
    \includegraphics[width=\linewidth]{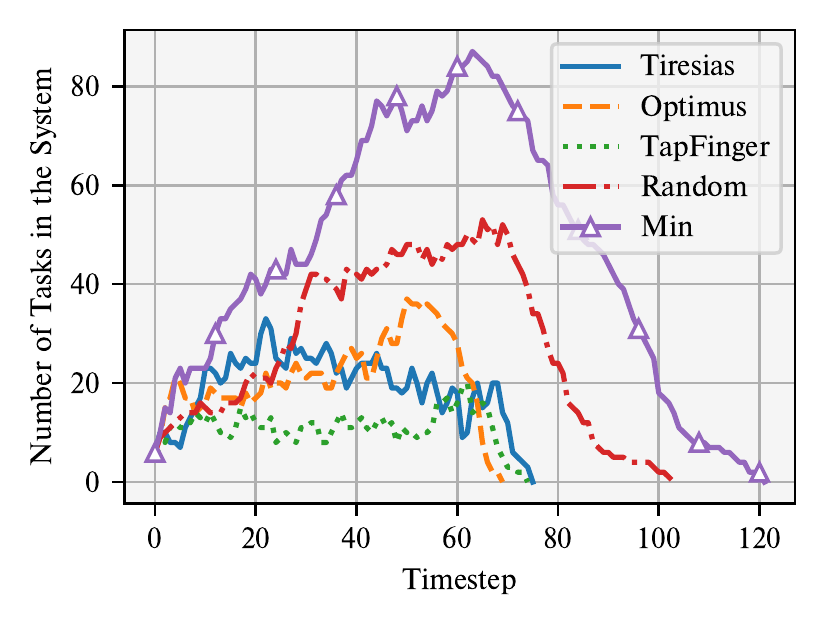}
    \caption{Task accumulation.}
    \label{fig:multiple_ta}
    \end{subfigure}\hfill
    \begin{subfigure}{0.5\linewidth}
    \includegraphics[width=\linewidth]{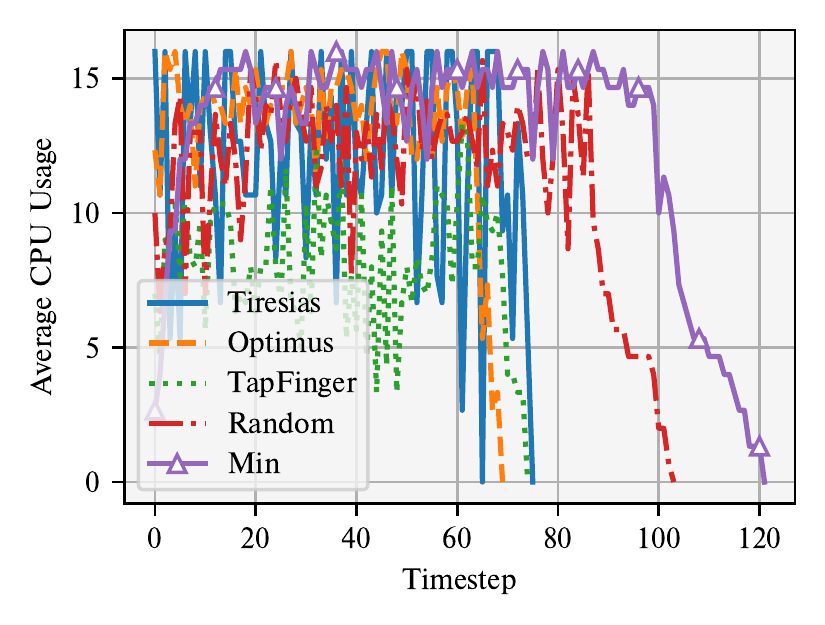}
    \caption{CPU utility.}
    \label{fig:multiple_cpu_utility}
    \end{subfigure}\hfill
    \begin{subfigure}{0.5\linewidth}
    \includegraphics[width=\linewidth]{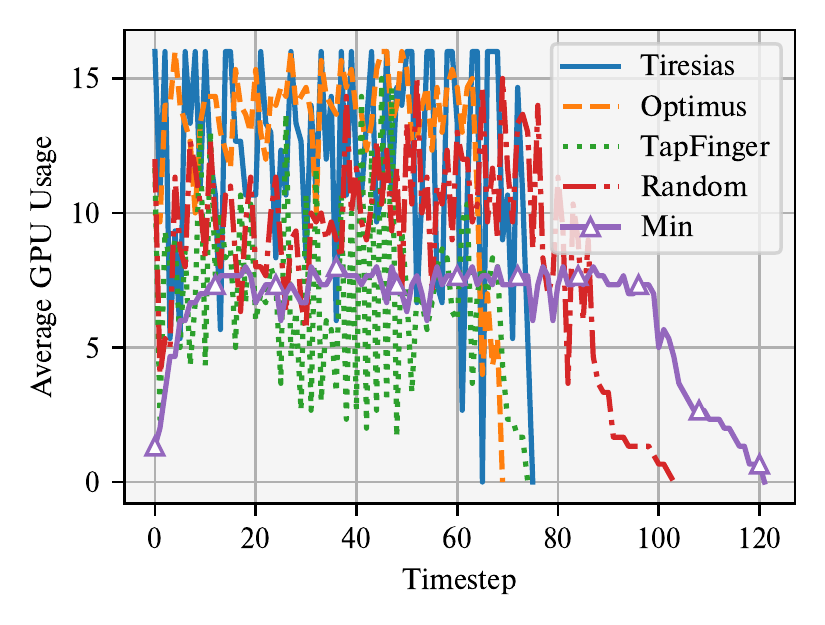}
    \caption{GPU utility.}
    \label{fig:multiple_gpu_utility}
    \end{subfigure}\hfill
    \caption{Multi-task {\em TapFinger} with 3-agent implementation: comparing different performance metrics with baselines.}
    \label{fig:multiple}
\end{figure}

{\bf Comparing the 3-agent implementation with baselines.} In the multi-task scheduling scenario, we 
grant no switching overhead to Tiresias and Optimus to demonstrate the superiority of the multi-task {\em TapFinger} over advanced preemptible rule-based algorithms.
As in Fig.s~\ref{fig:multiple_jct} and \ref{fig:multiple_ta}, Tiresias and Optimus are improved relative to themselves in the single-task settings and the other two heuristics because the switching overhead is ignored. However, the multi-task {\em TapFinger} still shows 43.2\% and 44.5\% reduction in terms of the average completion time, compared to Tiresias and Optimus, respectively.
The variance of the task completion times of the multi-task {\em TapFinger} is significantly reduced.
As the throughput of the multi-task {\em TapFinger} is greatly improved, the task accumulation is also mitigated to some extent.
We also look into the resource utilization of the multi-task {\em TapFinger} and other baselines without switching overhead caused by preemption. In Fig.s~\ref{fig:multiple_cpu_utility} and \ref{fig:multiple_gpu_utility}, the multi-task {\em TapFinger} shows considerable resource efficiency as it can achieve lower completion time with fewer resources.

\begin{figure}[!t]
    \centering
    \begin{subfigure}{0.5\linewidth}
    \includegraphics[width=\linewidth]{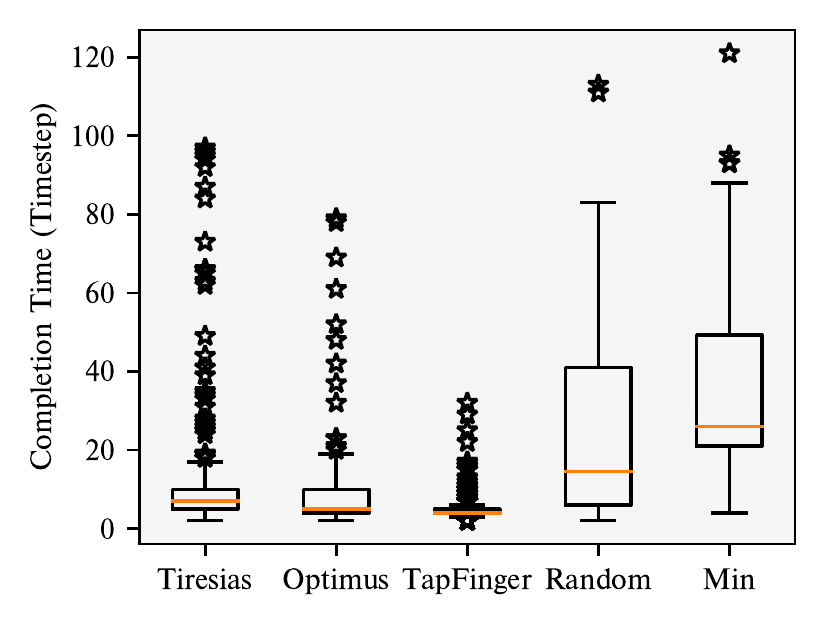}
    \caption{Completion time.}
    \label{fig:hetero_jct}
    \end{subfigure}\hfill
    \begin{subfigure}{0.5\linewidth}
    \includegraphics[width=\linewidth]{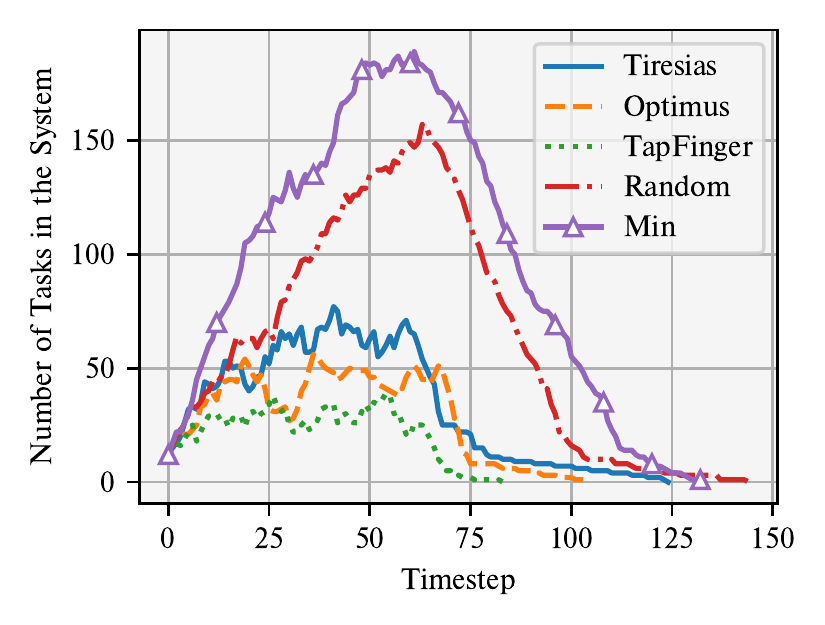}
    \caption{Task accumulation.}
    \label{fig:hetero_ta}
    \end{subfigure}\hfill
    \caption{Multi-task {\em TapFinger}: 6 heterogeneous edge clusters.}
    \label{fig:hetero}
\end{figure}

{\bf Heterogeneous clusters with 6-agent implementation.} We also verified that the multi-task {\em TapFinger} achieves good results on heterogeneous edge clusters. We raise the arrival rates of the same 6-agent synthetic workload in Section~\ref{subsubsec:synthetic} to 6, to stress the scheduling process. The resource capacities (number of CPU cores and GPUs) of the 6 edge clusters are $\{(4, 4), (8, 8), (12, 12), (16, 16), (24, 24), (32, 32)\}$.
Ideally, the clusters with larger resource capacity should take on heavier workloads to consume the shared resource-sensitive tasks in their queues. On the other hand, the clusters with limited resource capacities should schedule more resource-insensitive tasks. Otherwise, head-of-line blocking may occur due to the slow running of the resource-sensitive tasks with limited resources.
Fig.~\ref{fig:hetero_jct} shows great completion time improvements of the multi-task {\em TapFinger} over the best baselines, ranging from 40.1\% to 54.9\%.
In Fig.~\ref{fig:hetero_ta}, all the baselines exhibit a more severe long-tail effect on task accumulation than the multi-task {\em TapFinger}, which implies that the multi-task {\em TapFinger} is capable of achieving collaborative scheduling between heterogeneous agents. 

{\bf Comparing the multi-task and single-task versions of {\em TapFinger}.}
We evaluate the multi-task {\em TapFinger} based on the same synthetic workload settings as used to evaluate single-task {\em TapFinger}, with switching overhead disabled and multi-scheduling enabled. Comparing Fig.s~\ref{fig:multiple_jct}--\ref{fig:multiple_ta} with Fig.s~\ref{fig:synthetic_baseline_jct}--\ref{fig:synthetic_baseline_ta}, the multi-task {\em TapFinger} outperforms the single-task {\em TapFinger} with substantially lower completion time and task accumulation, which is due to the higher resource efficiency achieved by the multi-task scheduling.

\subsection{Ablation Analysis}
To better understand how different components of {\em TapFinger} affect the overall algorithm performance, we perform a comprehensive ablation analysis to investigate the individual contribution of each design choice for {\em TapFinger}. 
First, based on the multi-task {\em TapFinger} and the same 3-agent synthetic workload setting in Section~\ref{subsubsec:synthetic}, we replace the HAN by a 4-layer perceptron to encode the embedding of the pending tasks. To ensure the same information to be input to the new model, we concatenate the server information to every state entity of the pending tasks, i.e., concatenating $p_{j,n,t}$, $e_{n,t}$ and $q_{n,t}$. Other than that, all the other parts remain the same as the multi-task {\em TapFinger}.

\begin{figure}[!t]
    \centering
    \begin{subfigure}{0.5\linewidth}
    \includegraphics[width=\linewidth]{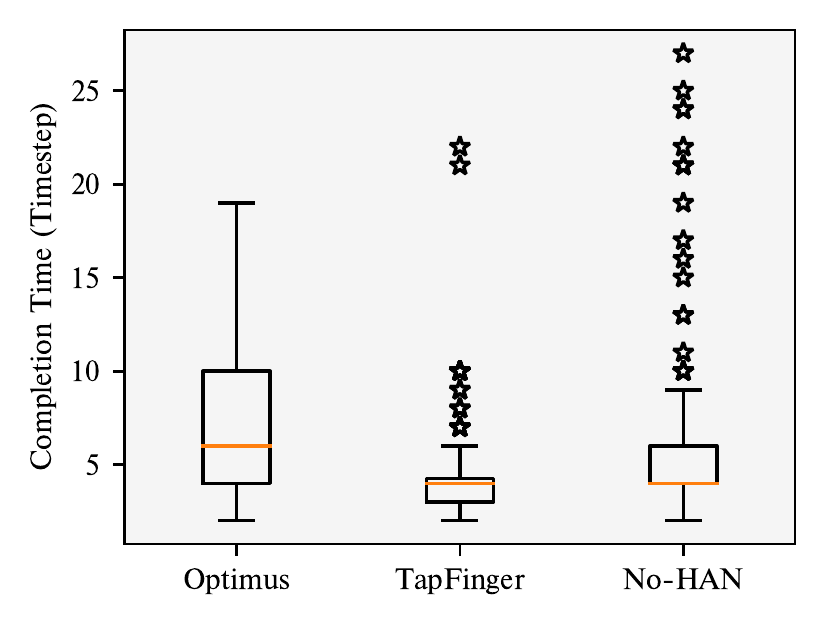}
    \caption{Completion time.}
    \label{fig:nognn_jct}
    \end{subfigure}\hfill
    \begin{subfigure}{0.5\linewidth}
    \includegraphics[width=\linewidth]{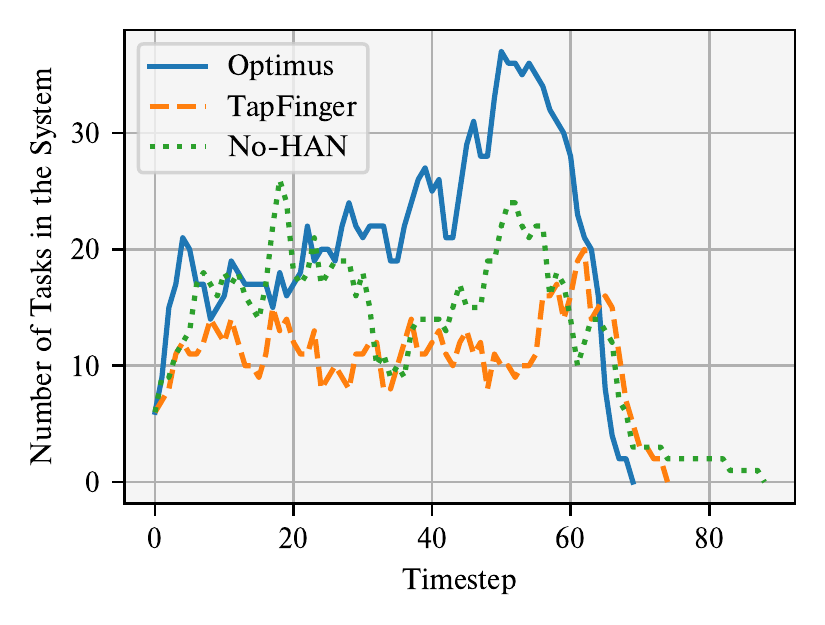}
    \caption{Task accumulation.}
    \label{fig:nognn_ta}
    \end{subfigure}\hfill
    \caption{Compared with the multi-task {\em TapFinger} without HAN on synthetic data.}
    \label{fig:nognn}
\end{figure}

{\bf Impact of state abstraction using HAN}. We first compare Optimus, our multi-task {\em TapFinger}, and the multi-task {\em TapFinger} without HAN (No-HAN). Fig.~\ref{fig:nognn_jct} shows that using multi-layer perceptron instead of HAN in {\em TapFinger} hurts performance, but is still better than Optimus.
Fig.~\ref{fig:nognn_ta} exhibits the emergence of a long-tail effect for the multi-task {\em TapFinger} without HAN, which implies that the ability of the agents to collaborate is diminished with HAN replaced.

We then switch to a new environment setting to highlight the effectiveness of the pointer mechanism. 
We consider 3 edge clusters and 2 resource types. The resource capacity and action dimension of each cluster are both 16 CPU cores and 16 GPUs.
The new workload has two task types. The minimum task resource requirement is 3 CPU cores and 3 GPUs for both tasks, with no progress gain for getting additional resources. One type of task lasts for 1 timestep, and the other lasts for 10 timesteps. For this kind of uncertain workload, the optimal scheduling policy should have subtle tradeoffs between resource efficiency and the preference for the shorter tasks\footnote{Shortest Job First is provably optimal in that it gives the lowest average completion time for a foreknown set of tasks.}.

\begin{figure}[!t]
    \centering
    \begin{subfigure}{0.5\linewidth}
    \includegraphics[width=\linewidth]{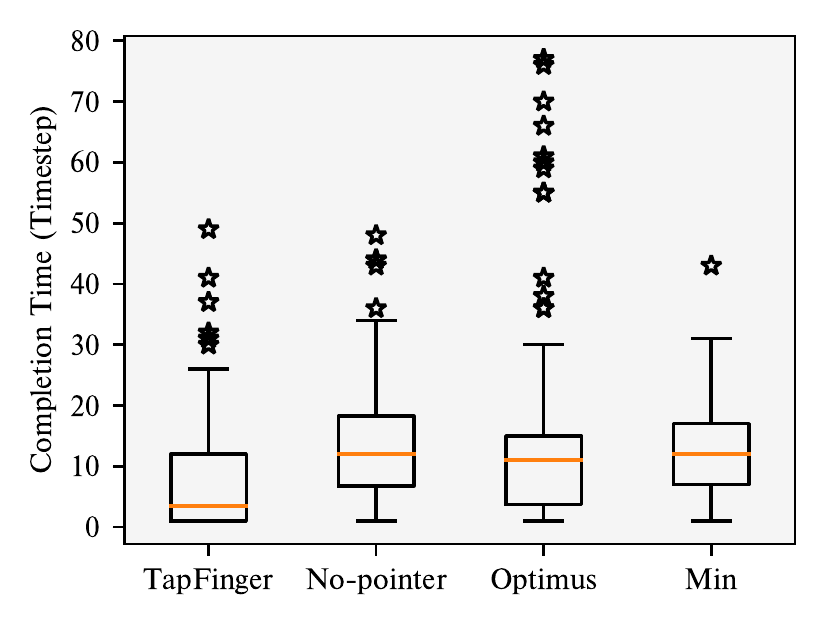}
    \caption{Completion time.}
    \label{fig:nopointer_jct}
    \end{subfigure}\hfill
    \begin{subfigure}{0.5\linewidth}
    \includegraphics[width=\linewidth]{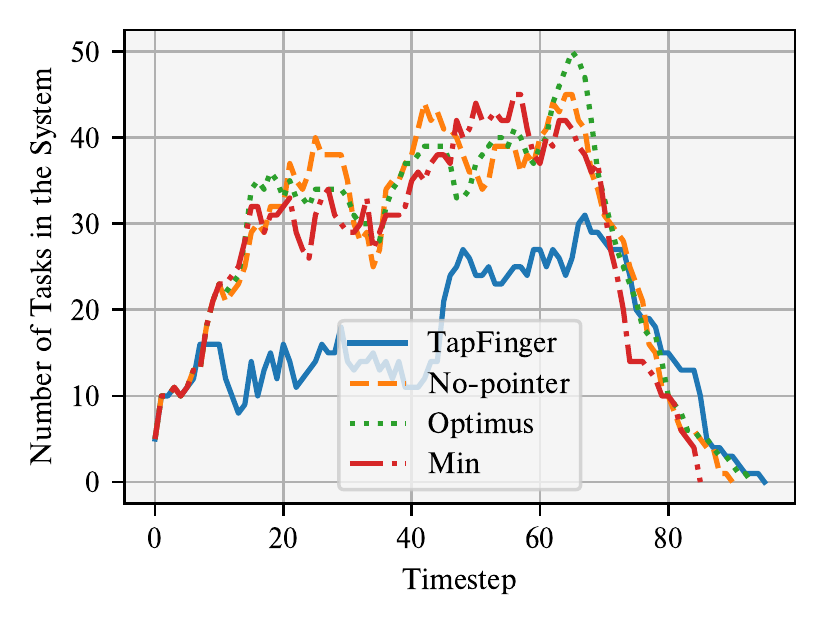}
    \caption{Task accumulation.}
    \label{fig:nopointer_ta}
    \end{subfigure}\hfill
    \caption{Compared with the multi-task {\em TapFinger} without pointer on synthetic data.}
    \label{fig:nopointer}
\end{figure}

{\bf Impact of augmenting actors using pointer networks.} We remove the pointer network from the multi-task {\em TapFinger}. Then, the multi-task {\em TapFinger} without the pointer mechanism randomly selects tasks from the pending set and then allocates resources to them using the same components as the original full {\em TapFinger}.
We compare the performance among Optimus, the multi-task {\em TapFinger}, the multi-task {\em TapFinger} without the pointer mechanism (named No-pointer), and a strategy that randomly selects tasks to schedule and allocates the minimum required amounts of resources to each task (named Min). Since there is no progress gain for additional resources allocated to the tasks, Min can achieve a completion time that is the lower-bound of any random task selection policies.
In Fig.s~\ref{fig:nopointer_jct} and \ref{fig:nopointer_ta}, the performance of the original multi-task {\em TapFinger}, which determines task selection by the pointer mechanism, is significantly better than other baselines in terms of the completion time and the system throughput.
We inspect several trajectories that the algorithms yield and find that the multi-task {\em TapFinger} with the pointer mechanism tends to schedule the longer tasks in the last when possible, but without additionally causing starvation, compared to the one with random task selection.

\section{Conclusion}\label{sec:conclusion}
We propose {\em TapFinger}, a distributed scheduling framework that jointly optimizes task placement and fine-grained multi-resource allocation for ML tasks in distributed edge clusters. 
{\em TapFinger} uses a MARL method based on HAN to encode the states of the edge components and their graph-structured 
interrelation semantics. We integrate the pointer network and the conflict resolution module into our actor network to decompose the actions. To mitigate the decision conflicts problem in MARL and to yield a valid resource allocation decision subject to complex constraints, we combine Bayes' theorem and masking schemes to construct the loss function for our model training. 
We provide both single-task and multi-task versions of {\em TapFinger} for different scenarios.
Our experiments show that {\em TapFinger} can reduce average completion times by up to 28.6\% and 14.5\% compared with the state-of-the-art scheduling algorithms on synthetic and test-bed ML workloads, respectively, and up to 54.9\% reduction using the extended multi-task {\em TapFinger}.

\section*{Acknowledgments}\label{sec:acknowledgments}
This work was supported by NSFC grants 62102460, 61902171, U1911201, U2001209, U20A20159, and 61972432, grants from Hong Kong RGC under the contracts HKU 17204619 and 17208920, grant 202201011392 under the Guangzhou Science and Technology Plan Project, and grant 2021B151520008 under the Guangdong Basic and Applied Basic Research Foundation.
\bibliographystyle{IEEEtran}
\bibliography{ref}
\begin{IEEEbiography}[{\includegraphics[width=1in,height=1.25in,clip,keepaspectratio]{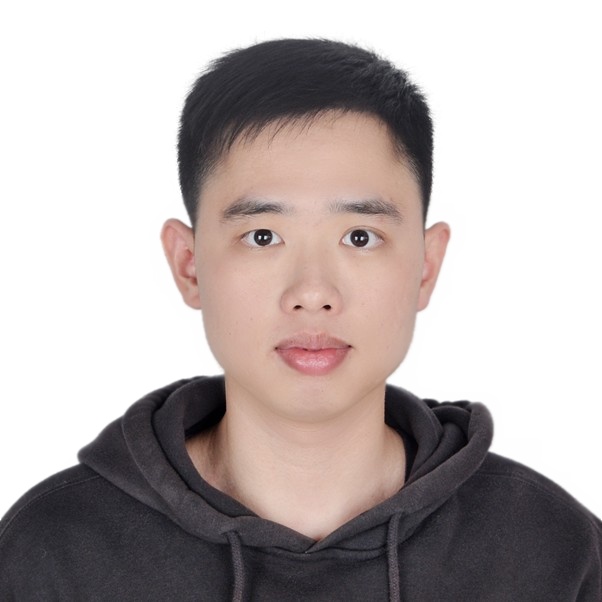}}]{Yihong Li}
received his bachelor’s degree from the School of Information Management, Sun Yat-sen University in 2021. He is currently pursuing a master’s degree with the School of Computer Science and Engineering, Sun Yat-sen University. His research interests include machine learning systems and networking.
\end{IEEEbiography}
\vskip -2\baselineskip plus -1fil
\begin{IEEEbiography}[{\includegraphics[width=1in,height=1.25in,clip,keepaspectratio]{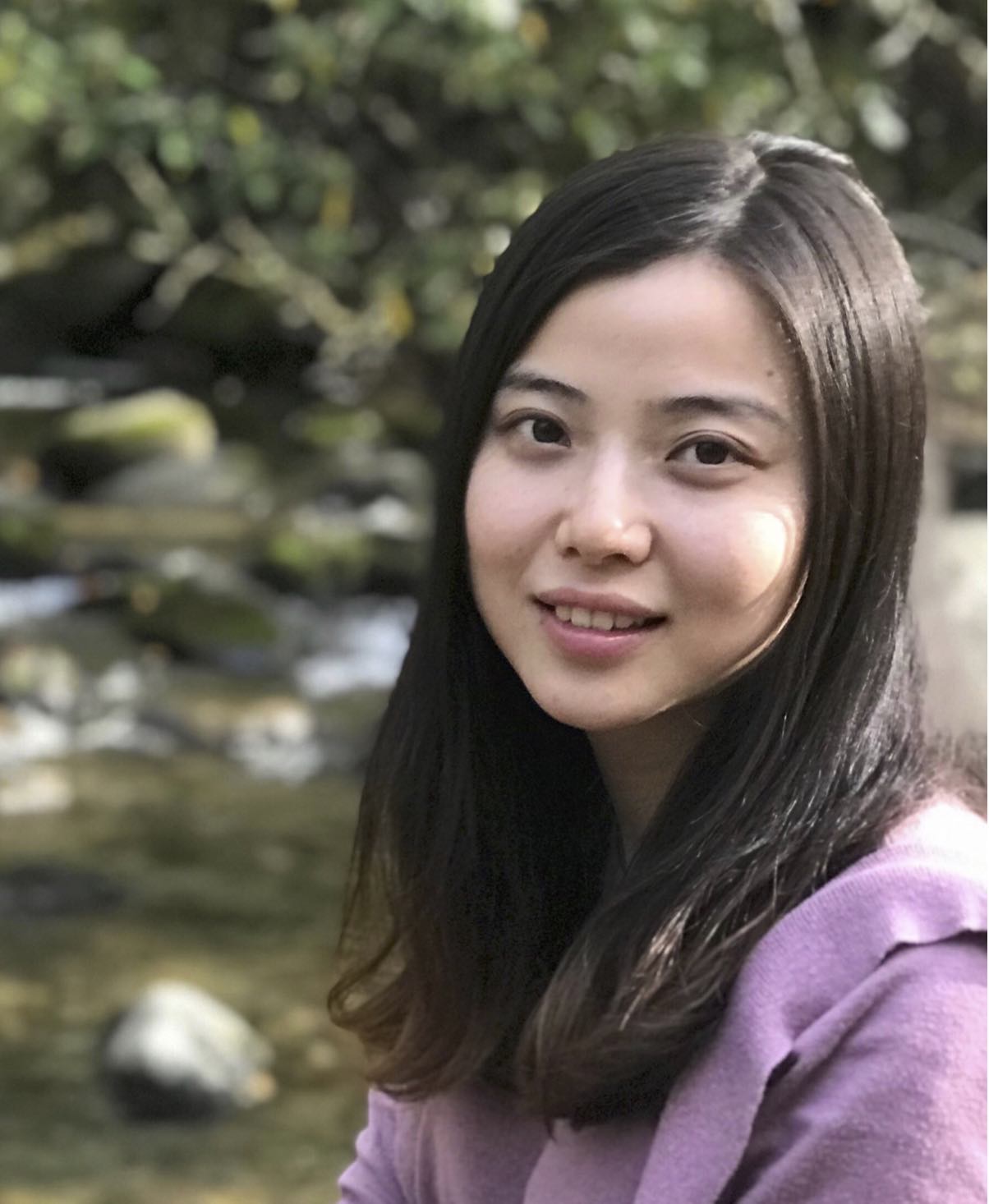}}]{Xiaoxi Zhang}
(Member, IEEE) received the B.E. degree in electronics and information engineering from the Huazhong University of Science and Technology in 2013 and the Ph.D. degree in computer science from The University of Hong Kong in 2017. She is currently an Associate Professor with the School of Computer Science and Engineering, Sun Yat-sen University. Before joining SYSU, she was a Post-Doctoral Researcher with the Department of Electrical and Computer Engineering, Carnegie Mellon University. She is broadly interested in optimization and algorithm design for networked systems, including cloud and edge computing networks, NFV systems, and distributed machine learning systems.
\end{IEEEbiography}
\vskip -2\baselineskip plus -1fil
\begin{IEEEbiography}[{\includegraphics[width=1in,height=1.25in,clip,keepaspectratio]{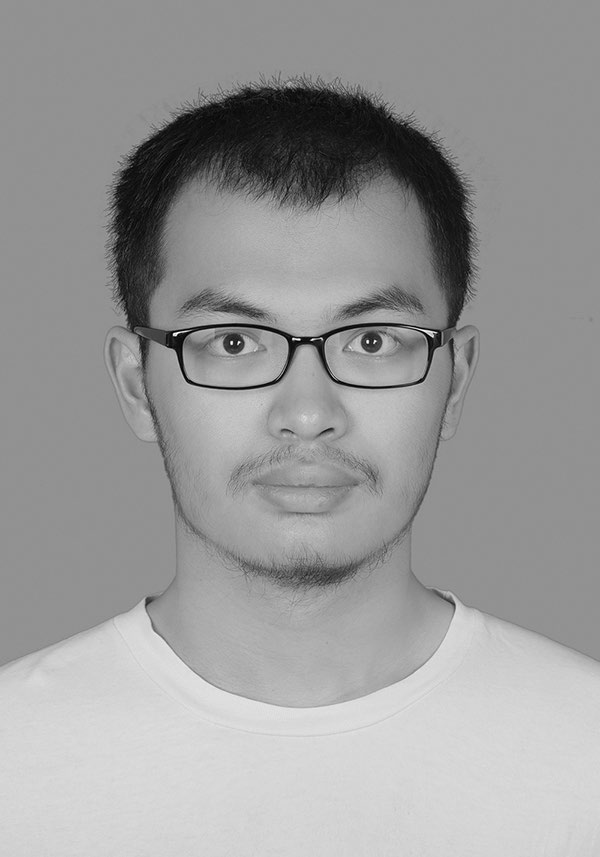}}]{Tianyu Zeng} 
 received the bachelor’s degree in 2021 from Sun Yat-sen University, Guangzhou, China, where he is currently working toward the master’s degree with the School of Computer Science and Engineering. His research interests include cloud and edge computing, distributed machine learning systems, and reinforcement learning.
\end{IEEEbiography}
\vskip -2\baselineskip plus -1fil
\begin{IEEEbiography}[{\includegraphics[width=1in,height=1.25in,clip,keepaspectratio]{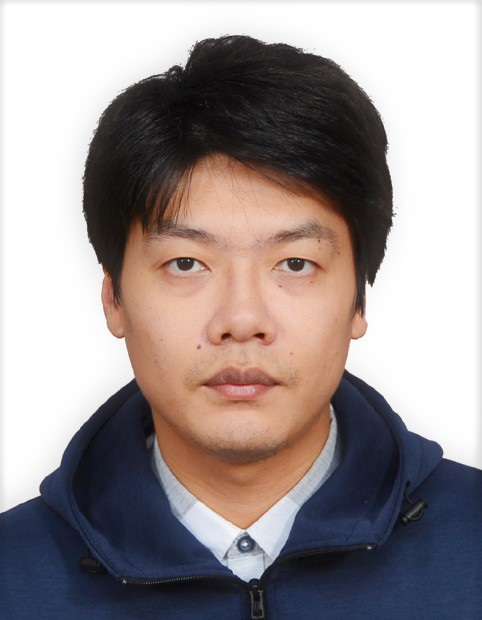}}]{Jingpu Duan}
(Member, IEEE) received the B.E. degree from the Huazhong University of Science and Technology, Wuhan, China, in 2013, and the Ph.D. degree from the University of Hong Kong, Hong Kong, China, in 2018. He is currently a Research Assistant Professor with the Institute of Future Networks, Southern University of Science and Technology, Shenzhen, China. He also works with the Department of Communications, Pengcheng Laboratory, Shenzhen, China. His research interest includes designing and implementing high-performance networking systems.
\end{IEEEbiography}
\vskip -2\baselineskip plus -1fil
\begin{IEEEbiography}[{\includegraphics[width=1in,height=1.25in,clip,keepaspectratio]{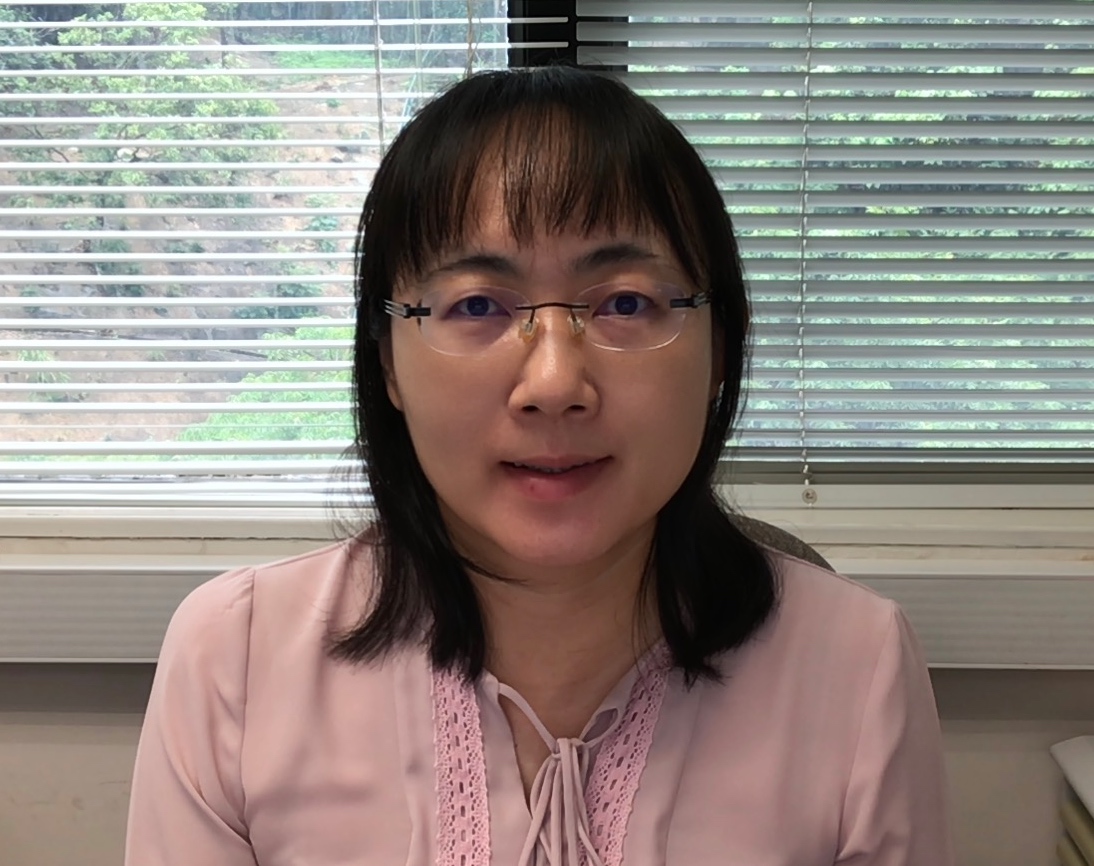}}]{Chuan Wu}
(Senior Member, IEEE) received the B.Eng. and M.Eng. degrees from the Department of Computer Science and Technology, Tsinghua University, China, in 2000 and 2002, respectively, and the Ph.D. degree from the Department of Electrical and Computer Engineering, University of Toronto, Canada, in 2008. Since September 2008, She has been with the Department of Computer Science, the University of Hong Kong, where she is currently a Professor. Her current research focuses on the areas of cloud computing, distributed machine learning systems and algorithms, and intelligent elderly care technologies. She is a member of ACM, and was the Chair of the Interest Group on Multimedia services and applications over Emerging Networks (MEN) of the IEEE Multimedia Communication Technical Committee (MMTC) from 2012 to 2014. She is an Associate Editor of IEEE TRANSACTIONS ON CLOUD COMPUTING, IEEE TRANSACTIONS ON MULTIMEDIA, ACM Transactions on Modeling and Performance Evaluation of Computing Systems, and IEEE TRANSACTIONS ON CIRCUITS AND SYSTEMS FOR VIDEO TECHNOLOGY. She was the co-recipient of the best paper awards of HotPOST 2012 and ACM e-Energy 2016.
\end{IEEEbiography}
\vskip -2\baselineskip plus -1fil
\begin{IEEEbiography}[{\includegraphics[width=1in,height=1.25in,clip,keepaspectratio]{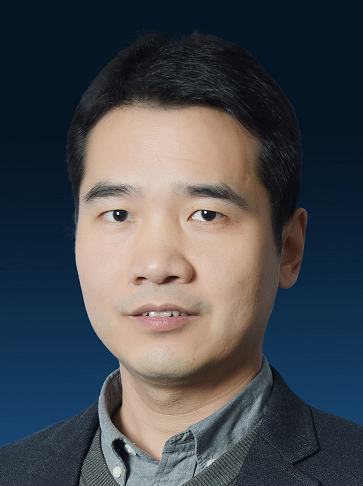}}]{Di Wu}
(M’06-SM’17) received the B.S. degree from the University of Science and Technology of China, Hefei, China, in 2000, the M.S. degree from the Institute of Computing Technology, Chinese Academy of Sciences, Beijing, China, in 2003, and the Ph.D. degree in computer science and engineering from the Chinese University of Hong Kong, Hong Kong, in 2007. He was a PostDoctoral Researcher with the Department of Computer Science and Engineering, Polytechnic Institute of New York University, Brooklyn, NY, USA, from 2007 to 2009, advised by Prof. K. W. Ross. Dr. Wu is currently a Professor and the Associate Dean of the School of Computer Science and Engineering with Sun Yat-sen University, Guangzhou, China. He was the recipient of the IEEE INFOCOM 2009 Best Paper Award, IEEE Jack Neubauer Memorial Award, and etc. His research interests include edge/cloud computing, multimedia communication, Internet measurement, and network security.

\end{IEEEbiography}
\vskip -2\baselineskip plus -1fil
\begin{IEEEbiography}[{\includegraphics[width=1in,height=1.25in,clip,keepaspectratio]{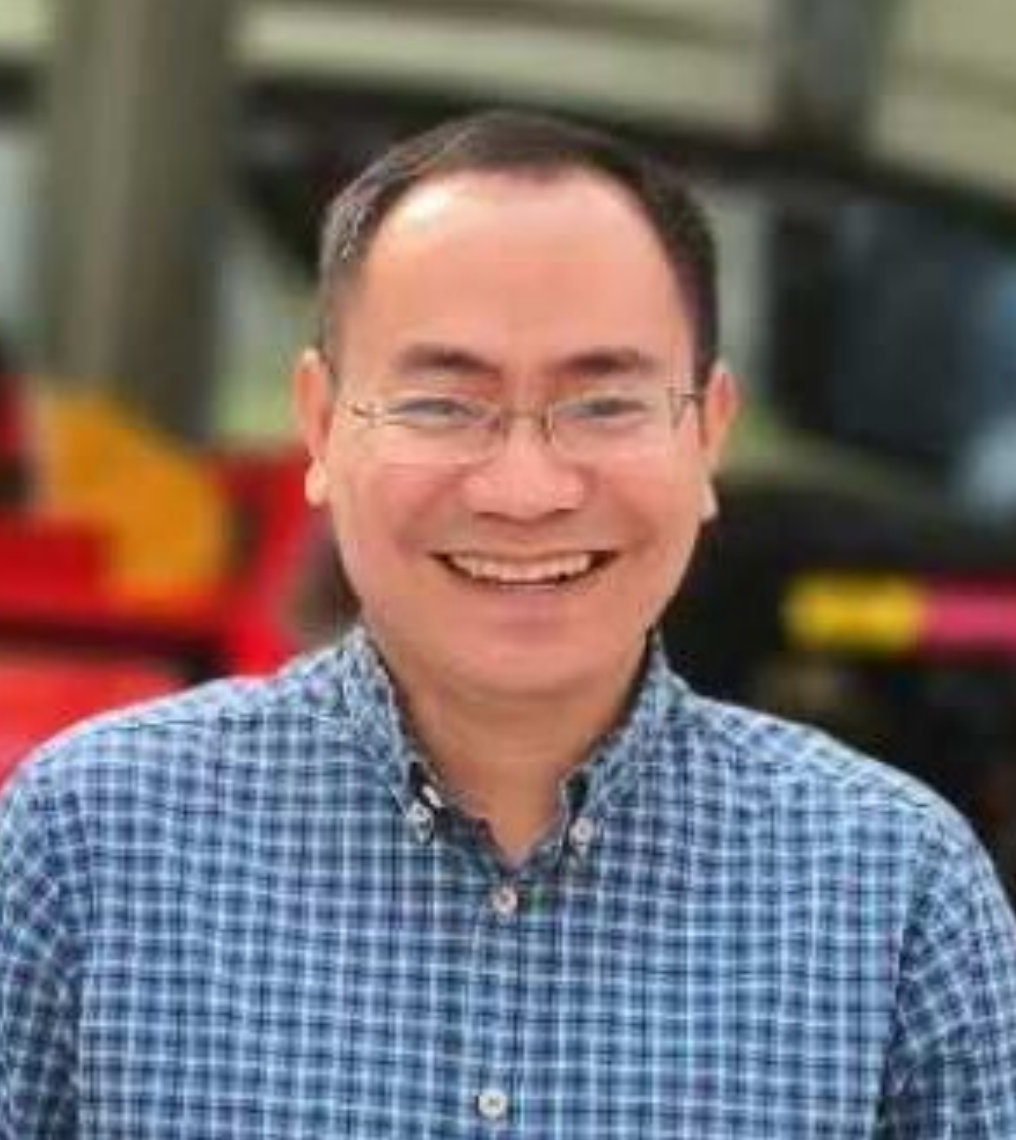}}]{Chen Xu}
is a Full Professor with Sun Yat-sen University, Guangzhou, China, and the Vice Director of National and Local Joint Engineering Laboratory of Digital Home Interactive Applications. He received the Ph.D. degree in information engineering from the Chinese University of Hong Kong in 2012, and worked as a Postdoctoral Research Associate at Arizona State University, Tempe, USA, from 2012 to 2014, an d a Humboldt Scholar Fellow at the Institute of Computer Science of the University of Goettingen, Germany from 2014 to 2016. He received the prestigious Humboldt research fellowship awarded by the Alexander von Humboldt Foundation of Germany, 2014 Hong Kong Young Scientist Runner-up Award, 2017 IEEE Communication Society Asia-Pacific Outstanding Young Researcher Award, 2017 IEEE ComSoc Young Professional Best Paper Award, Honorable Mention Award of 2010 IEEE international conference on Intelligence and Security Informatics (ISI), Best Paper Runner-up Award of 2014 IEEE International Conference on Computer Communications (INFOCOM), and Best Paper Award of 2017 IEEE Intranational Conference on Communications (ICC). He is currently an Area Editor of the IEEE OPEN JOURNAL OF THE Communications Society, an Associate Editor of the IEEE TRANSACTIONS WIRELESS COMMUNICATIONS, IEEE INTERNET OF THINGS JOURNAL and IEEE JOURNAL ON SELECTED AREAS IN COMMUNICATIONS (JSAC) Series on Network Softwarization and Enablers.
\end{IEEEbiography}

\end{document}